\documentclass[12pt]{article}
\pdfoutput=1
\usepackage{subfigure}
\usepackage{amssymb,amsmath}
\usepackage{graphicx}
\usepackage{setspace,caption}
\usepackage{lipsum}
\usepackage[normalem]{ulem}

\usepackage{color}
\usepackage[colorlinks=true
,urlcolor=blue
,citecolor=blue
,linkcolor=blue
,pagecolor=blue
,linktocpage=true
,pdfproducer=medialab
]{hyperref}
\usepackage[a4paper]{geometry}

\usepackage{placeins}
\usepackage{cite}
\usepackage{cancel}

\makeatletter \renewcommand{\@dotsep}{10000} \makeatother

\mathchardef\mhyphen="2D

\newcommand{\beq}{\begin{equation}}
\newcommand{\eeq}{\end{equation}}
\newcommand{\bea}{\begin{eqnarray}}
\newcommand{\eea}{\end{eqnarray}}

\begin{document}

\begin{titlepage}
\pagestyle{empty}


\vspace*{0.2in}
\begin{center}
	\baselineskip 20pt 
	{\Large\bf  LHC Run-3, $b-\tau$ Yukawa Unification and Dark Matter Implications in SUSY 4-2-2 model}
	
	\vspace{1cm}

	{\large 
		Waqas Ahmed$^{a}$\footnote{E-mail: \texttt{\href{mailto:waqasmit@hbpu.edu.cn}{waqasmit@hbpu.edu.cn}}},
		Mohamed Belfkir$^{b}$\footnote{E-mail: \texttt{\href{mailto:mohamed.belfkir@cern.ch}{mohamed.belfkir@cern.ch}}},
		Salah Nasri$^{b, c}$\footnote{E-mail: \texttt{\href{mailto:snasri@uaeu.ac.ae; salah.nasri@cern.ch}{snasri@uaeu.ac.ae; salah.nasri@cern.ch}}},
		Shabbar Raza$^{d}$\footnote{E-mail: \texttt{\href{mailto:shabbar.raza@fuuast.edu.pk}{shabbar.raza@fuuast.edu.pk}}} and 
		Umer Zubair$^{e}$\footnote{E-mail: \texttt{\href{mailto:umer@udel.edu}{umer@udel.edu}}} 
		
	} 
	\vspace{.5cm}
	
	{\baselineskip 20pt \it
		$^{a}$ \it
		School of Mathematics and Physics, Hubei Polytechnic University, \\
		Huangshi 435003,
		China \\
		\vspace*{6pt}
		$^{b}$Department of Physics, United Arab Emirates University,\\
		Al Ain 15551 Abu Dhabi, UAE\\
		\vspace*{6pt}
		$^{c}$International Center for Theoretical Physics, Trieste, Italy\\
		\vspace*{6pt}
		$^{d}$Department of Physics, Federal Urdu University of Arts, Science and Technology, Karachi 75300, Pakistan \\
		\vspace*{6pt}
		$^{e}$Division of Science and Engineering,  \\ Pennsylvania State University, Abington, PA 19001, USA \\
		
		\vspace{2mm} }
\end{center}

\begin{abstract}
\noindent

\end{abstract}
We revisit the bottom and $\tau$ Yukawa coupling unification in supersymmetric $4$-$2$-$2$ model and present for the first time the sbottom-neutralino co-annihilation scenario consistent with the the bottom and $\tau$ Yukawa coupling unification. In addition, we show gluino-neutralino, stop-neutralino, stau-neutralino, chargino-neutralino and A-resonance scenario and show that all such solutions are consistent with existing experimental collider constraints, Planck2018 dark matter relic density bounds as well as direct and indirect bounds on neutralino-nucleons scattering cross sections. We show that in sbottom-neutralino co-annihilation scenario, the sbottom mass is about 2 TeV whereas in the case of gluino-neutralino, stop-neutralino, the gluino mass can be between 1 TeV to 3 TeV and stop mass in the range of 1 TeV to 3.5 TeV. {Moreover, in the case of co-annihilation scenario, the stau and chargino masses can be as heavy as 3.5 TeV,}  while the A-resonance solutions are in the range of 0.5 TeV to 3.5 TeV. We anticipate that some part of the parameter space will be  accessible in the supersymmetry searches at LHC Run-3 and beyond.
\end{titlepage}



\section{Introduction}
\label{sec:intro}
The beauty of Supersymmetric Standard Models (SUSY SMs) is that they provide: gauge coupling unification ~\cite{gaugeunification}, solution to gauge hierarchy problem \cite{ghp} and a candidate dark matter particle if augmented with $R$ parity conservation \cite{Jungman:1995df}. It should be noted that the Minimal Supersymmetric Standard Model (MSSM) predicts Higgs boson with mass $m_{h}\lesssim$ 135 GeV \cite{mhiggs} whereas, the Higgs mass observed at the Large Hadron collider (LHC) is about 125 GeV \cite{Aad:2012tfa,CMS}. Moreover, another unification which models like SUSY $SO(10)$ and SUSY $SU(4)_c\times SU(2)_{L}\times SU(2)_{R}$ (4-2-2) can accommodate is the unification of top ($t$), bottom ($b$) and tau ($\tau$) Yukawa unification (YU) $t$-$b$-$\tau$ or ($b$-$\tau$) \cite{big-422,bigger-422,Baer:2008jn,Gogoladze:2009ug,Baer:2009ff,Gogoladze:2009bn,Gogoladze:2010fu} (For latest study in SUSY and non-SUSY frameworks, see \cite{Gomez:2020gav,Djouadi:2022gws}). In SUSY 4-2-2 model, the soft supersymmetry breaking (SSB) mass terms for gauginos $M_{1}$, $M_{2}$ and $M_{3}$ corresponding to $U(1)_{Y}$, $SU(2)_{L}$ and $SU(3)_{c}$, respectively can be given as 
\begin{equation}
M_1=\frac{3}{5}M_2+\frac{2}{5}M_3.
\end{equation}
This non-universality of guaginos along with the sign of Higgsino mass parameter $\mu$ can be utilized to explore very interesting phenomenology of SUSY 4-2-2. Ref. \cite{Gogoladze:2010fu} showed the importance of $\mu <0 $ in achieving correct threshold corrections to the bottom quark Yukawa coupling and having light spectrum consistent with $t$-$b$-$\tau$ YU. It should be noted that SUSY 4-2-2 is the only model that yields gluino-neutralino co-annihilation solutions consistent with dark matter relic density and 10$\%$ or better $t$-$b$-$\tau$ YU \cite{Gogoladze:2009ug, Gogoladze:2009bn,Profumo:2004wk,Ajaib1}. 

It was also shown in refs. \cite{Gogoladze:2009ug,Gogoladze:2009bn,Profumo:2004wk} that $t$-$b$-$\tau$ YU in $4$-$2$-$2$ with the same sign SSB gaugino mass terms is consistent with LSP neutralino dark matter through gluino-neutralino coannihilation channel. Moreover, {\color{red}for} the combination $\mu <0$ and gauginos with $M_{2} < 0$ and $M_{3} > 0$, it is shown that solutions consistent with experimental constraints along with 10$\%$ or better $t$-$b$-$\tau$ YU can be realized in $4$-$2$-$2$ for $m_{0} \gtrsim 300$ GeV, as opposed to $m_{0} \gtrsim 8$ TeV for the case of same sign gaugino masses, where $m_{0}$ represents the universal SSB mass parameter for scalars at  $M_{GUT}$ \cite{Gogoladze:2010fu}. In this case, co-annihilation scenarios such as chargino-neutralino and stau-neutralino are available along with A-funnel channel to achieve the correct dark relic density \cite{Gogoladze:2010fu,Gogoladze:2012ii}. 

It should be noted that in general, $t$-$b$-$\tau$ YU is maintained in 4-2-2 but not necessarily be kept intact if higher dimensional operators are also considered.  In such a scenario, one can consider a set of higher dimensional operators whose contributions to the Yukawa couplings are expressed as $y_{e}/y_{d}=$1 and $y_{u}/y_{d} \neq$ 1 \cite{Antusch:2013rxa,Antusch:2009gu,Trine:2009ns} such that one can still maintain $b$-$\tau$ YU in 4-2-2 but not $t$-$b$-$\tau$ YU.

In this article, we update the status of $b$-$\tau$ YU in line of the work reported in ref. \cite{Raza:2014upa} in the light of LHC Run-3 and new SUSY searches. In ref. \cite{Raza:2014upa}, $t$-$b$-$\tau$ YU and $b$-$\tau$ YU are considered with same sign gaugino mass parameters and $\mu > 0$. In this study{\color{red},} it is shown that the co-annihilation of  Next to Lightest SUSY particle (NLSP) gluino, with the LSP neutralino, is the only channel available to obtain solutions consistent with experimental bounds and dark matter relic density bounds consistent with 10$\%$ or better $t$-$b$-$\tau$ YU. It should be noted that in such a scenario the heaviest NLSP gluino mass reported was about 1 TeV. In $b$-$\tau$ YU case, light stop NLSP co-annihilation scenario with LSP neutralino  was shown besides gluino-neutralino co-annihilation. In this case NLSP gluino mass still remained around 1 TeV but the NLSP light stop mass was around 0.8 TeV. In a recent study of $t$-$b$-$\tau$ YU in SUSY 4-2-2 with $\mu >0$ but non-universal scalar mass parameters and gauginos with relative signs{\color{red},} see \cite{Gomez:2020gav}, where it is shown that there exit NLSP gluino, NLSP stau, NLSP chargino co-annihilation with LSP neutralino and A-resonance solutions satisfying  experimental constraints along with dark matter relic density bounds and $R_{tb\tau}\lesssim$1.1. In this article, we employ relative sign gaugino mass parameters and $\mu <0$ and study sparticle spectrum consistent with collider bounds, 10$\%$ or better $b$-$\tau$ YU and dark matter relic density constraints in SUSY 4-2-2 framework. Since $b$-$\tau$ YU is a relaxed constraint as compared to $t$-$b$-$\tau$ YU, we expect more richer phenomenology. In fact we do have very interesting phenomenological scenarios. For the first time we report the NLSP sbottom co-annihilation with LSP neutralino scenario in SUSY 4-2-2 consistent with $b$-$\tau$ YU \footnote{In fact a couple of NLSP sbottom solutions also satisfy $t$-$b$-$\tau$ YU within 5$\%$. In this article we are reporting the NLSP sbottom scenario and detailed study of such a scenario will be presented elsewhere \cite{self}.}. To the best of our knowledge, ref, \cite{Gogoladze:2011ug} is the only paper which discusses sbottom-neutralino co-annihilation in SUSY $SU(5)$ framework. In refs. \cite{Gogoladze:2011ug,Baer:2012by} it was shown that sbottom-neutralino co-annihilation solutions consistent with experimental bounds were not compatible with $b$-$\tau$ YU in $SU(5)$. In fact the NSP sbottom co-annihilation with the  LSP neutralino requires non-trivial relationship among the SSB parameter. Besides, sbottom-neutralino co-annihilation, we also have gluino-neutralino, stop-neutralino, stau-neutralino, chargino-neutralino and A(H)-resonance solutions compatible with 10$\%$ or better $b$-$\tau$ YU and consistent with present experimental constraints. We show that our solutions are compatible with recent LHC SUSY searches, LHC Run-3 and future projections. In addition, our solutions also satisfy dark matter constraints, such as Planck2018 dark matter relic density bounds, dark matter direct and indirect current and future bounds.

The fundamental parameters of the 4-2-2 model under consideration are give as:
\begin{align}
m_{0}, m_{H_u}, m_{H_d}, A_0, M_2, M_3,  \tan\beta, {\rm sign}(\mu).
\label{params}
\end{align}
Here $m_0$ is the universal SSB mass for MSSM sfermions, $m_{H_{u,d}}$ are Higgs SSB mass terms, $A_{0}$ is the universal teri-linear scalar couplings, $M_2$ and $M_{3}$, as discussed before, are the gauginos SSB mass terms. All these parameters are defined at $M_{GUT}$. The parameter $\tan\beta \equiv v_{u}/v_{d} $, which is the ratio of the vacuum expectation values (VEVs) of the two MSSM Higgs doublets, is defined at low scale.                                                                                                                                              

The outline for the remainder of this paper is as follows. The summary of the scanning procedure and the experimental constraints employed in our analysis is given in section \ref{sec:scan}. We show results of scans for $b$-$\tau$ YU in section \ref{sec:results}. We also provide a table of six benchmark points as an example of our results. Our conclusion is summarized in section \ref{sec:conc}.


\section{Scanning Procedure and Experimental Constraints}
\label{sec:scan}
We use the ISAJET 7.85 package \cite{ISAJET} to perform random scans on model parameters. In ISAJET,  the unification condition is $g_U=g_1=g_2$ at $M_{\rm GUT}$, and allow $g_3$ to deviate within 3$\%$. We assign such such deviation due to unknown threshold corrections at the GUT scale \cite{Hisano:1992jj}. For a details discussion on the ISAJET package working, see ref \cite{ISAJET,bartol2}.

The fundamental parameters defined earlier are chosen in the following ranges:
\begin{gather}
0 ~ \text{TeV} \leq  m_{0}, m_{H_u}, m_{H_d} \leq 20 ~ \rm{TeV} \nonumber \\
-10 ~ \text{TeV} \leq M_2  \leq 10 ~ \text{TeV} \nonumber \\
0 ~ \text{TeV} \leq M_3  \leq 5 ~ \text{TeV} \nonumber \\
30 \leq \tan\beta \leq 55 \nonumber \\
-3 \leq A_{0}/m_0 \leq 3 \nonumber \\
\mu < 0.
\label{parameterRange}
\end{gather}

we employ the Metropolis-Hastings algorithm in scanning the parameter space \cite{Belanger:2009ti}. We collect only those points which have successful radiative electroweak symmetry breaking (REWSB) and neutralino is the LSP in this way we exclude solutions where charged particles are stable \cite{Beringer:1900zz}. A part from these conditions, we also impose the mass bounds on all the sparticles \cite{Agashe:2014kda}, and the constraints from rare decay processes; $B_{s}\rightarrow \mu^{+}\mu^{-} $ \cite{Aaij:2012nna}, $b\rightarrow s \gamma$ \cite{Amhis:2012bh}, and $B_{u}\rightarrow \tau\nu_{\tau}$ \cite{Asner:2010qj}. We also required LHC constraints on gluino and first/second generation squark masses \cite{Vami:2019slp} as well as the relic abundance of the LSP neutralino to satisfy the $5\sigma$ bounds of Planck 2018 data \cite{Ade:2015xua}. More explicitly, we set;
\begin{gather}
m_h  = (122 - 128) ~ {\rm GeV} \\
m_{\tilde{g}} \geq 2.3 ~ {\rm TeV}, \quad m_{\tilde{q}} \geq 2 ~ {\rm TeV}\\
0.8\times 10^{-9} \leq{\rm BR}(B_s \rightarrow \mu^+ \mu^-) \leq 6.2 \times10^{-9} \;(2\sigma) \\
2.99 \times 10^{-4} \leq {\rm BR} (b \rightarrow s \gamma) \leq 3.87 \times 10^{-4} \; (2\sigma) \\
0.15 \leq \frac{{\rm BR}(B_u\rightarrow\tau \nu_{\tau})_{\rm MSSM}} {{\rm BR}(B_u\rightarrow \tau \nu_{\tau})_{\rm SM}} \leq 2.41 \; (3\sigma)\\
0.114 \leq \Omega_{\rm CDM}h^2 (\rm Planck~2018) \leq 0.126   \; (5\sigma).
\end{gather}
Apart from these constraints, we quantify $b-\tau$ YU with the parameter $R_{b\tau}$ defined as \cite{Belanger:2009ti}
\begin{equation}
R_{tb\tau}\equiv \dfrac{{\rm max}(y_{b},y_{\tau})}{{\rm min}(y_{b},y_{\tau})},
\end{equation}
where $R_{b\tau}=1$ implies perfect $b-\tau$ YU. However, we allow 10$\%$ ($R_{b\tau}=1.1$) variation from the perfect unification due to various uncertainties.

\section{Results}
\label{sec:results}
\subsection{Fundamental Parameter Space for $b-\tau$ YU}
\label{sec:btauYU}

\begin{figure}[h!]
\centering \includegraphics[width=8cm]{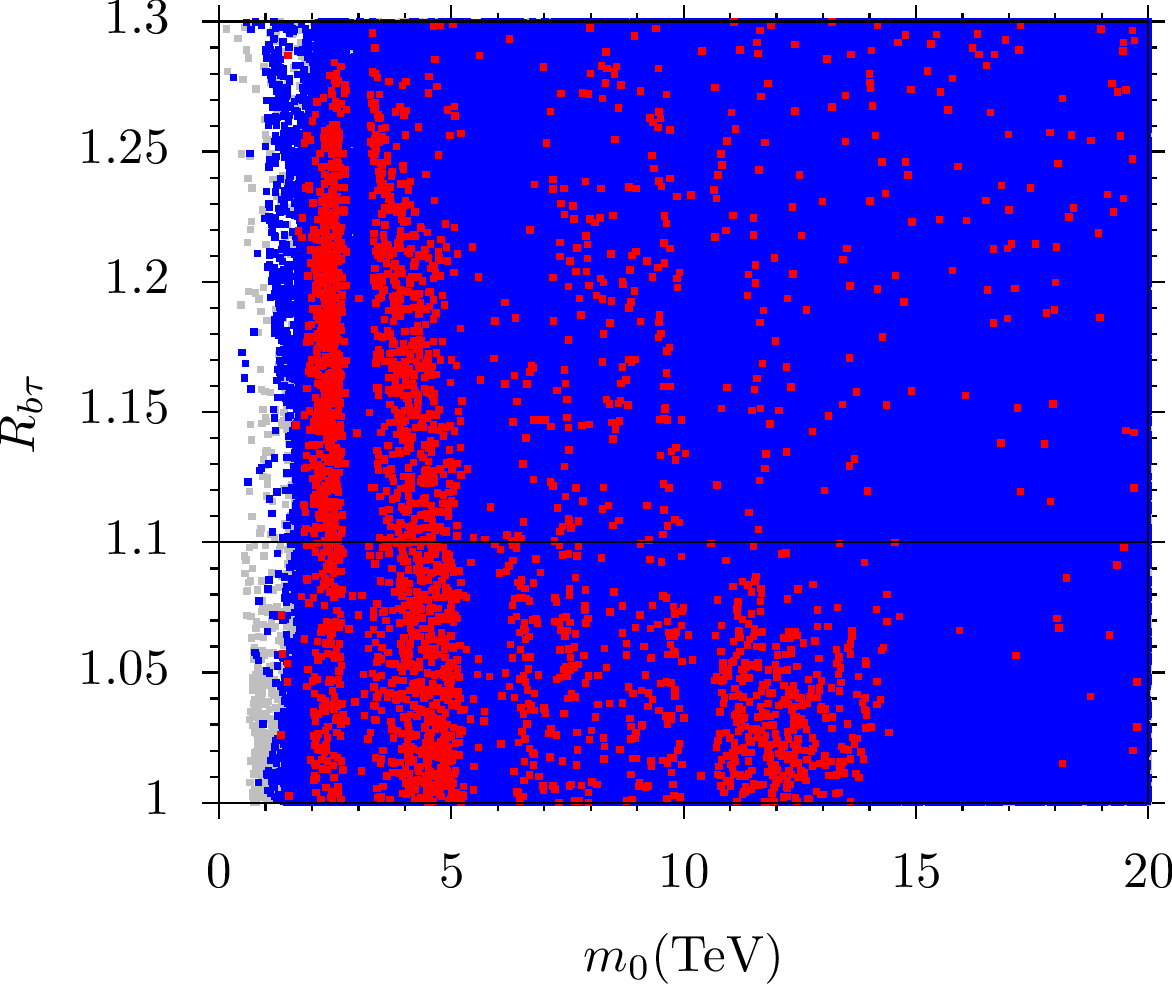}
\centering \includegraphics[width=8cm]{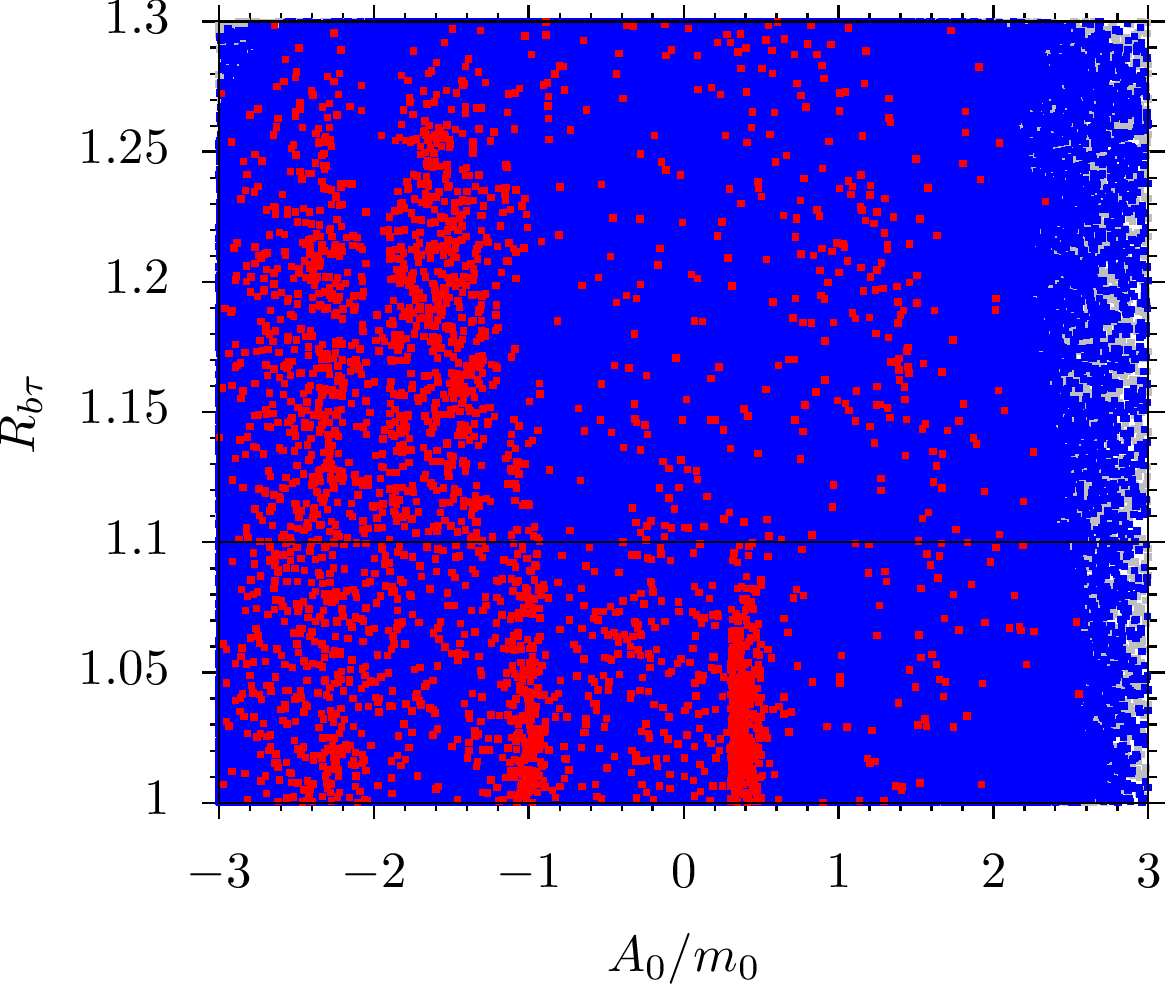}
\centering \includegraphics[width=8cm]{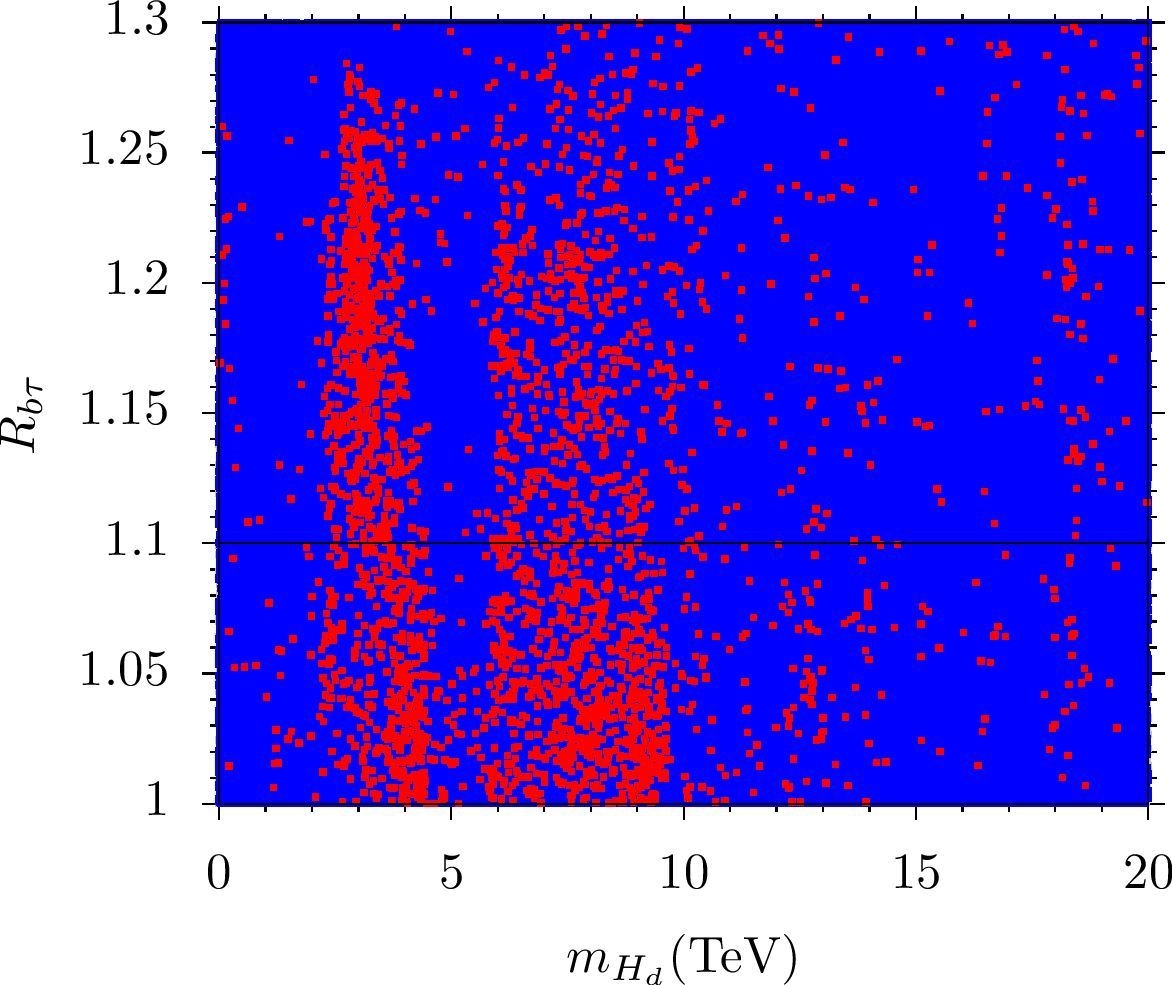}
\centering \includegraphics[width=8cm]{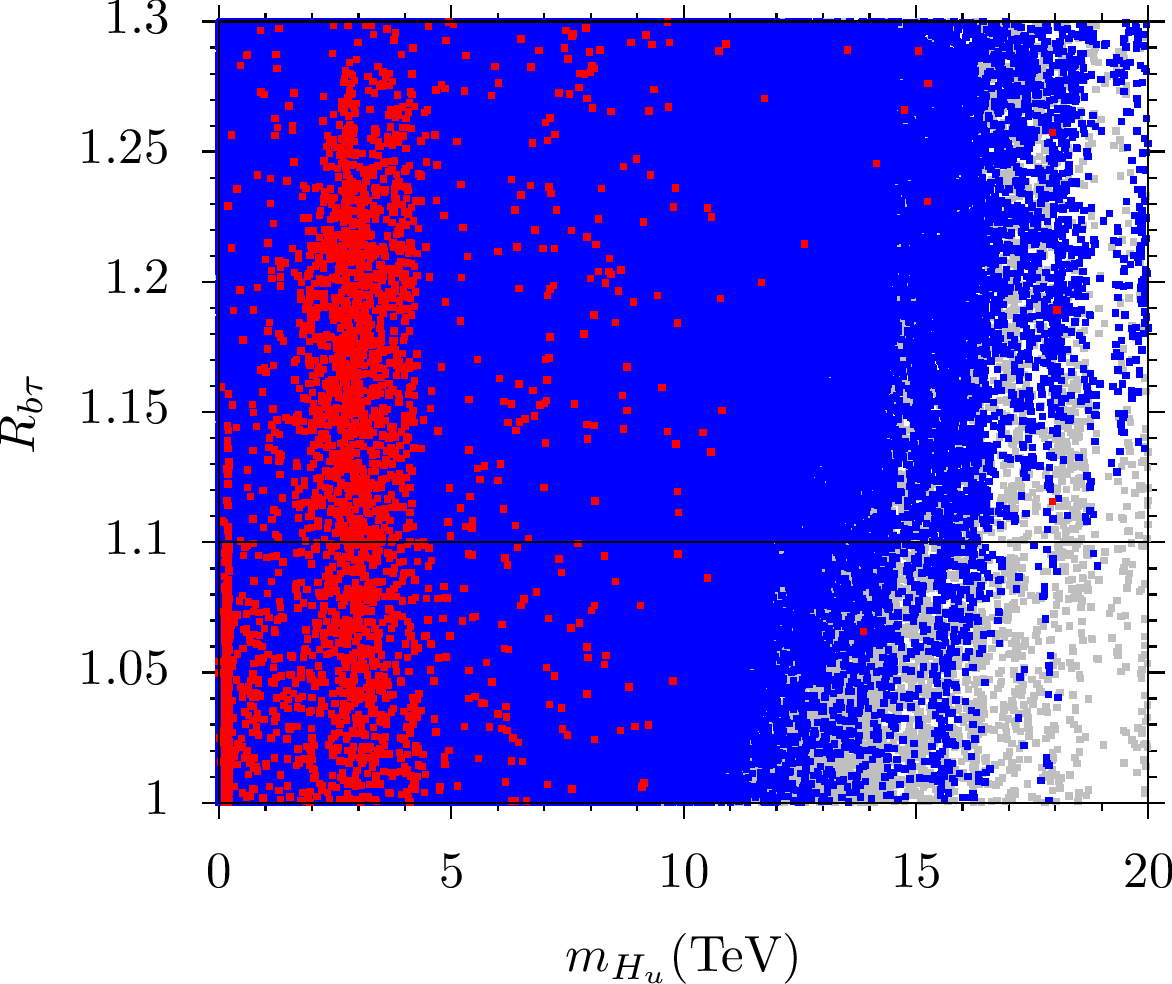}
\caption{Plots in the $m_{0}-R_{b\tau}$, ${A_0/m_0}-R_{b\tau}$,  $m_{H_d}-R_{b\tau}$ and $m_{H_u}-R_{b\tau}$ planes. Gray points are consistent with the REWSB and LSP neutralino LSP conditions. Blue points represent sparticle mass bounds, Higgs mass bound, B-physics bounds and  red points points satisfy $5\sigma$ Planck2018 bounds on the relic density of the LSP neutralino. The horizontal line shows the regions with $R_{b\tau}=1.1$, below which are the solutions with 10$\%$ or better $b-\tau$ YU.}
\label{fig1}
\end{figure}
\begin{figure}[h!]
\centering \includegraphics[width=8cm]{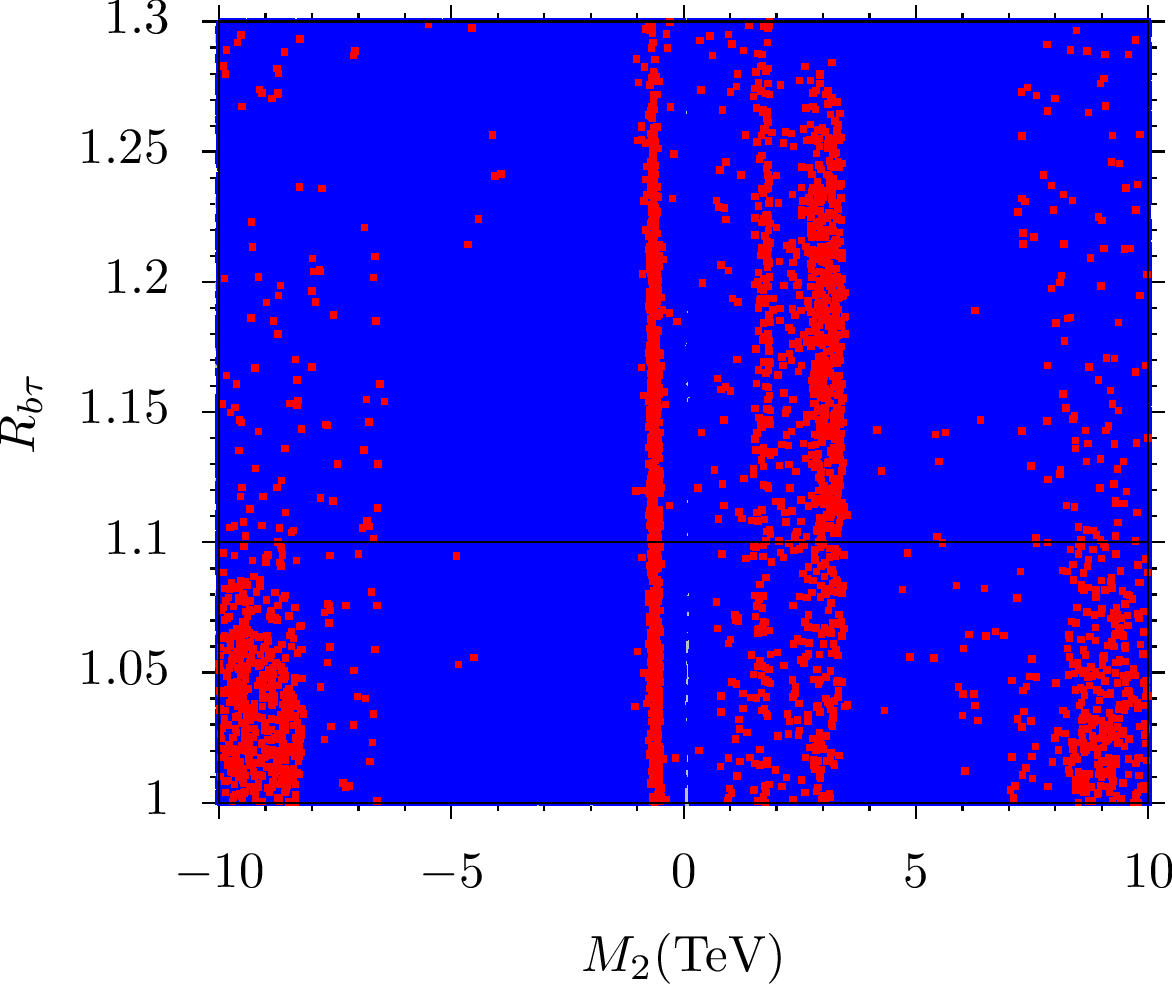}
\centering \includegraphics[width=8cm]{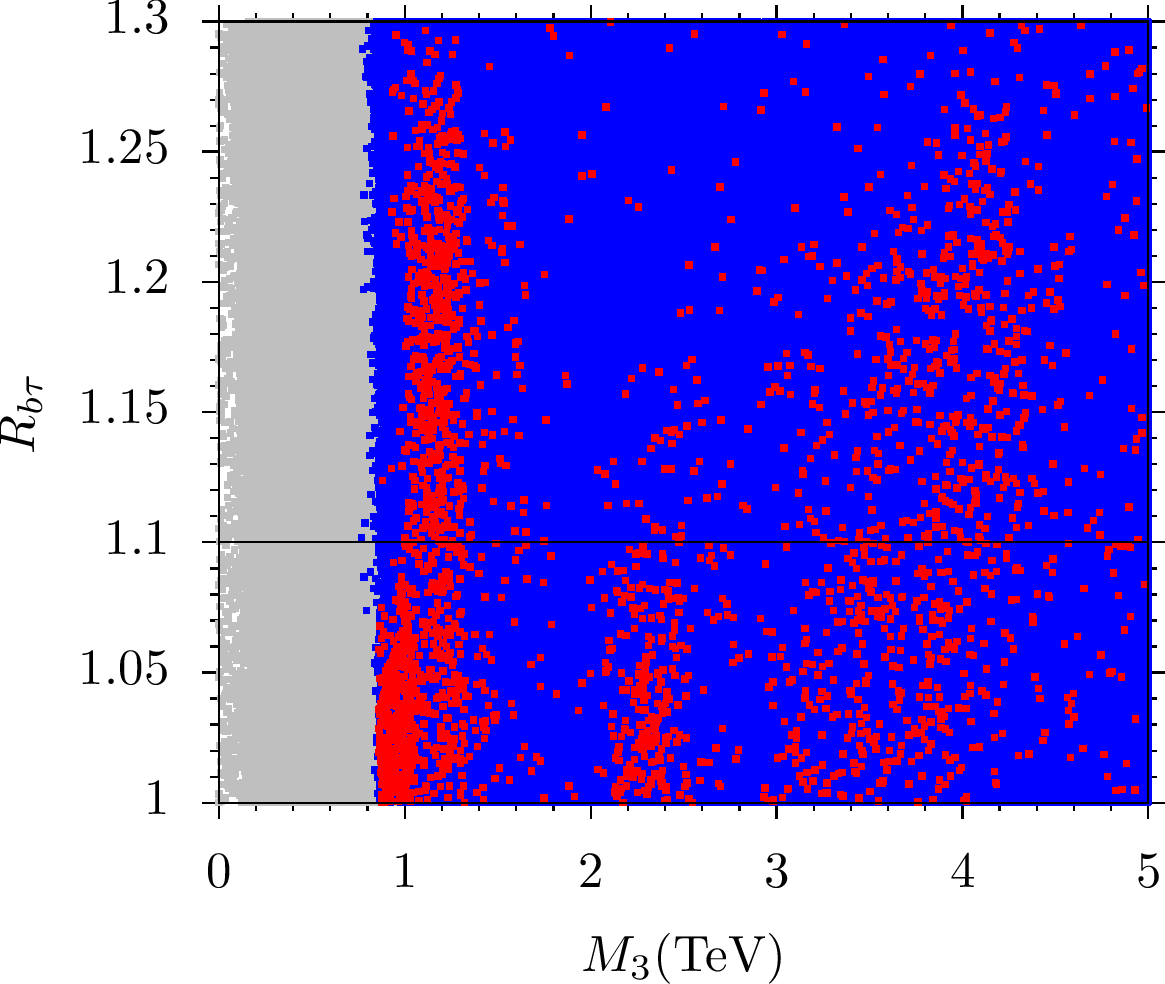}
\centering \includegraphics[width=8cm]{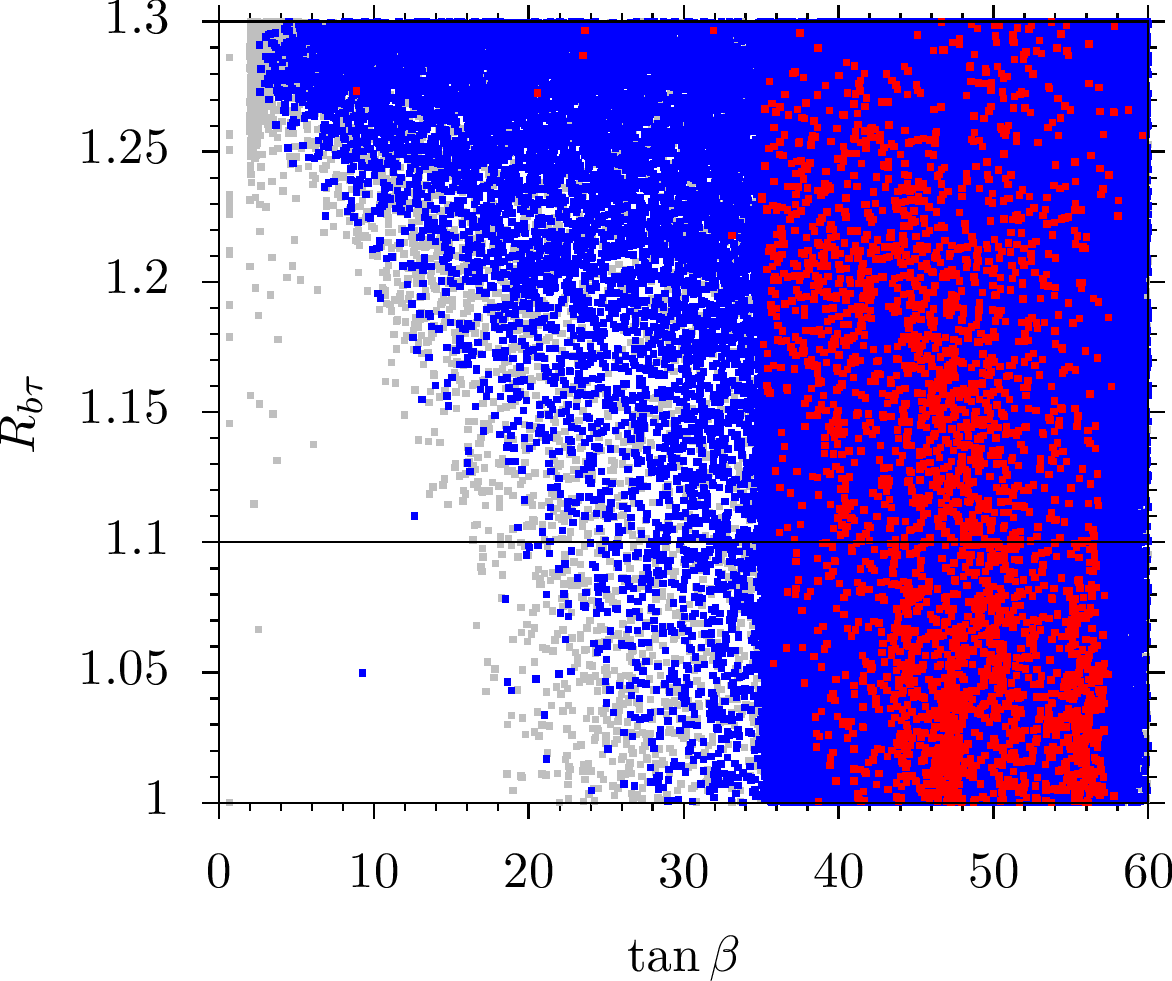}
\caption{Plots in the $ M_{2}-R_{b\tau}$ and $M_{3}-R_{b\tau}$ and $\tan\beta-R_{b\tau}$, planes. The color coding is the same as in Figure \ref{fig1}.}
\label{fig2}
\end{figure}
\begin{figure}[h!]
\centering \includegraphics[width=8cm]{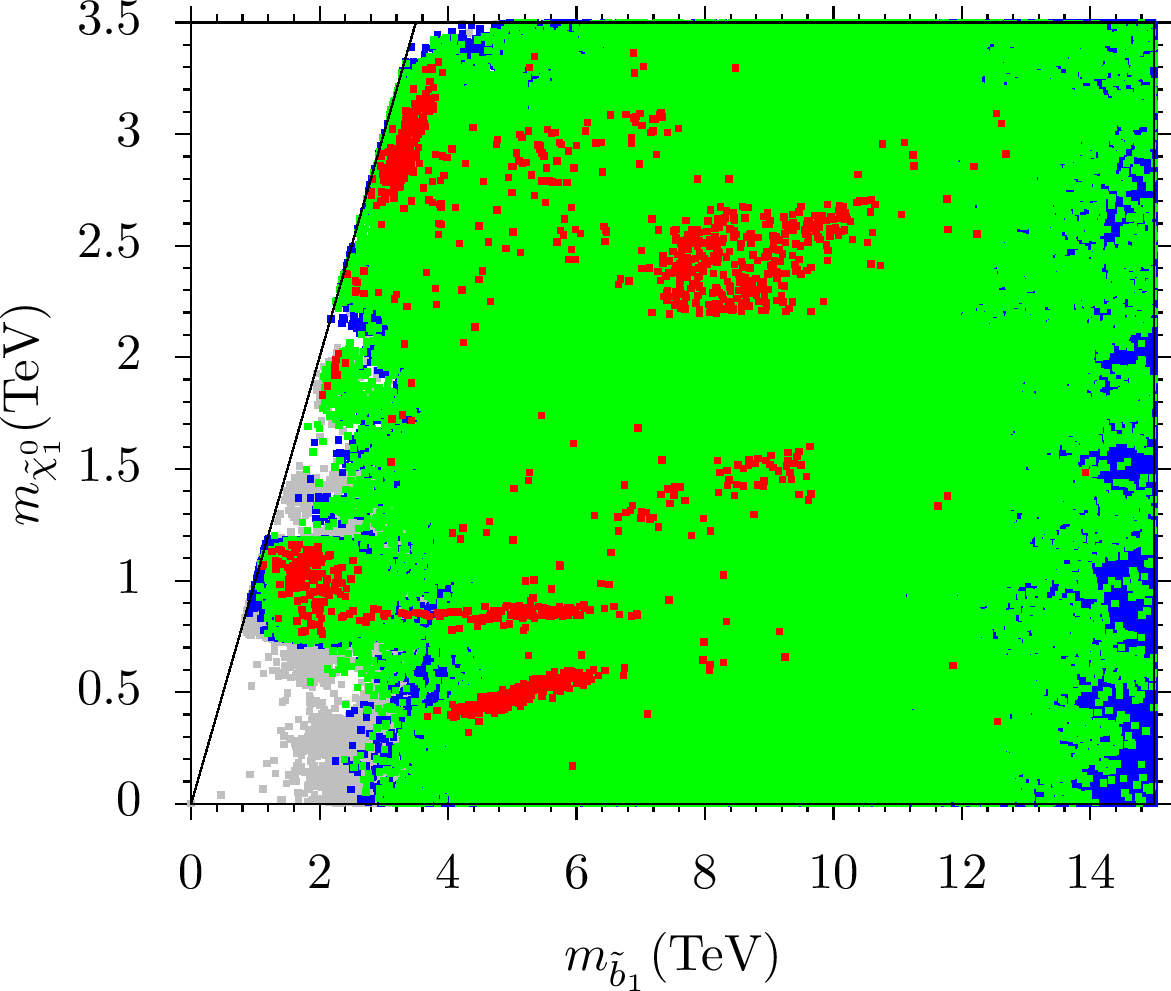}
\centering \includegraphics[width=8cm]{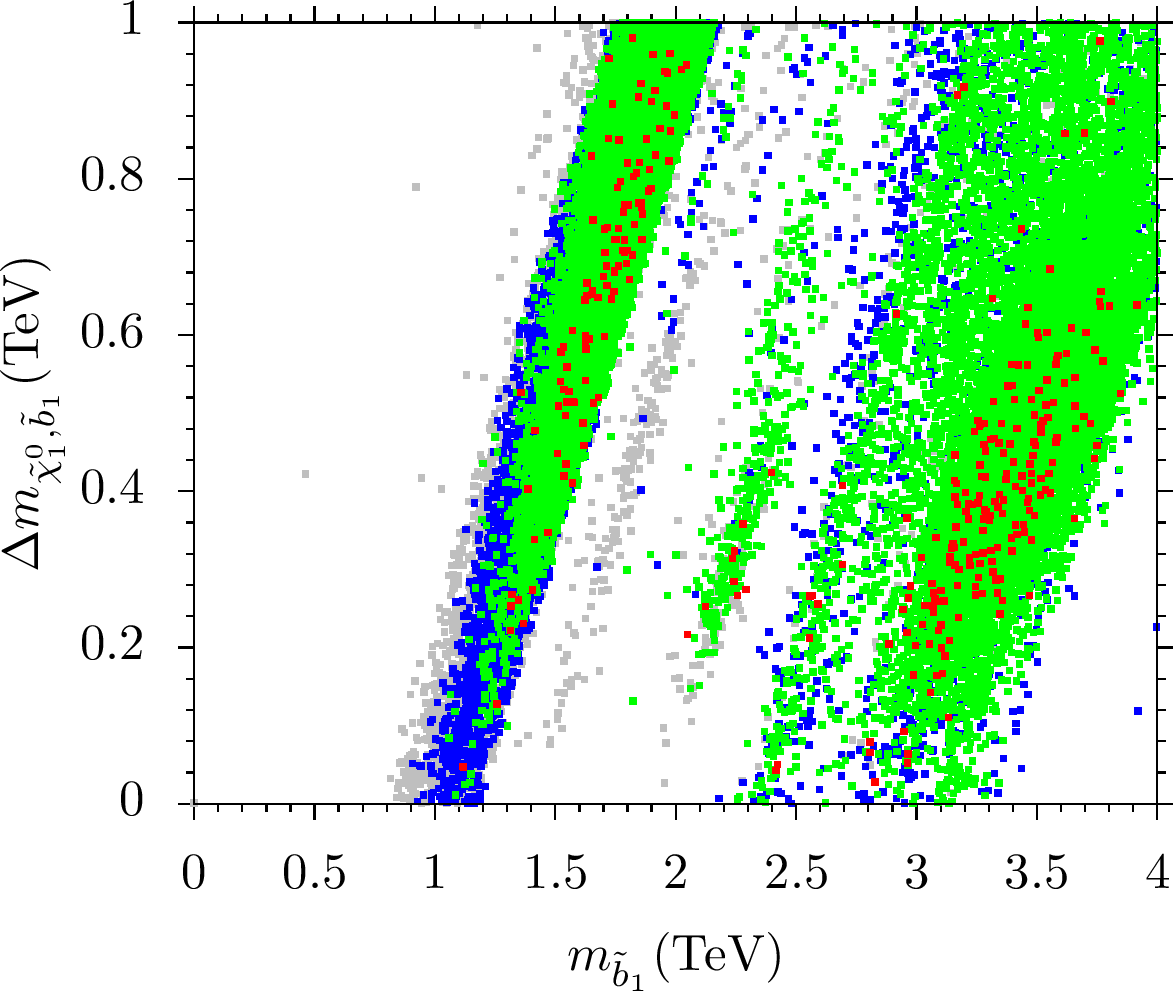}
\caption{Plots in the $m_{\tilde b_{1}}-m_{\tilde \chi_{1}^{0}}$ and $m_{\tilde b_{1}}-\mid \Delta m_{\tilde \chi_{1}^{0},\tilde b_{1}}\mid$ planes. Gray points are compatible with the REWSB and LSP neutralino conditions. Blue points represent sparticle mass bounds, Higgs mass bound, B-physics bounds. Green points form subset of blue points and have $R_{b\tau}\lesssim 1.1$. Red points are a subset of green points, and they satisfy $5\sigma$ the Planck2018 bound on the relic density of the LSP neutralino.}
\label{fig3}
\end{figure}
\begin{figure}[h!]
\centering \includegraphics[width=8cm]{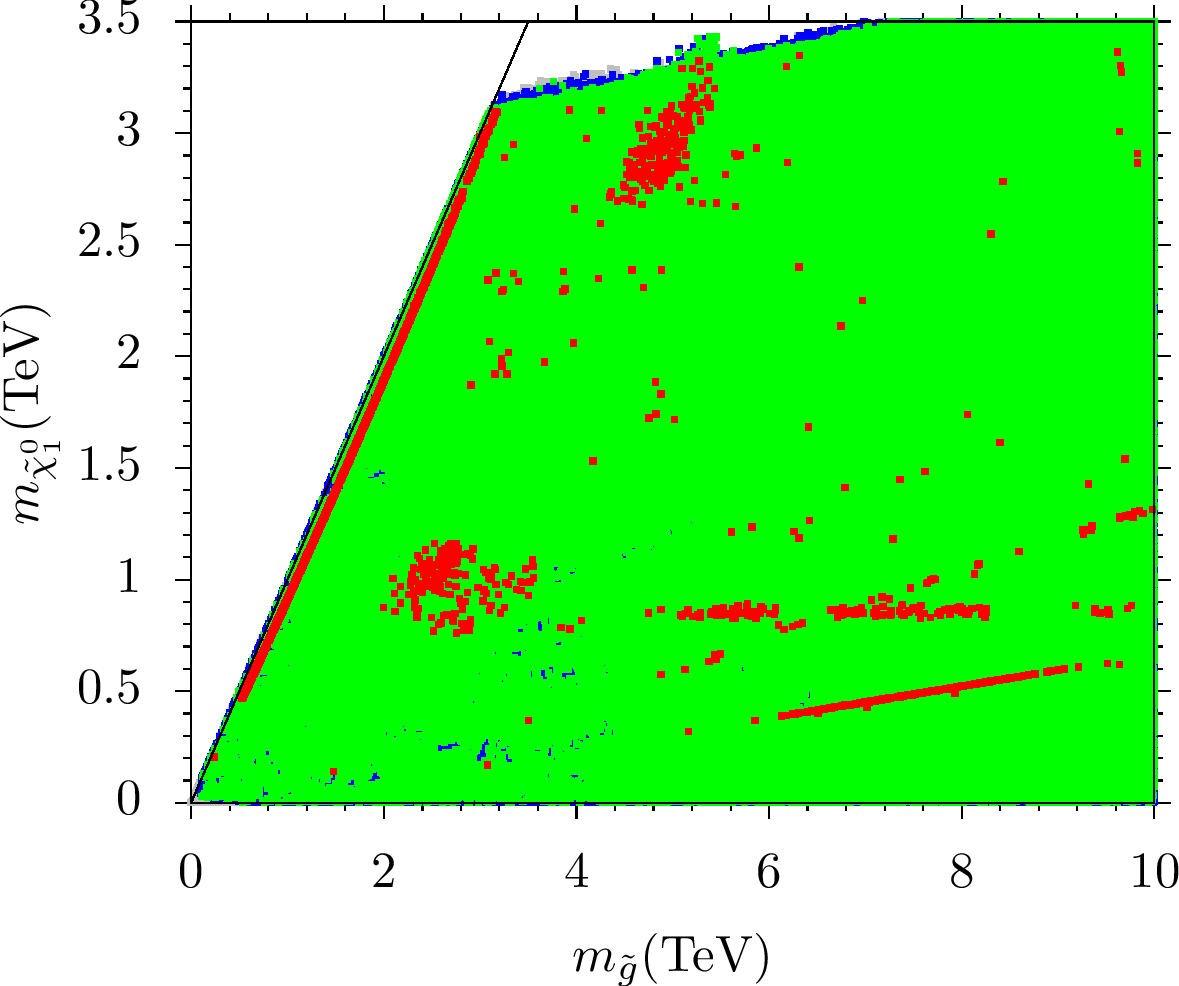}
\centering \includegraphics[width=8cm]{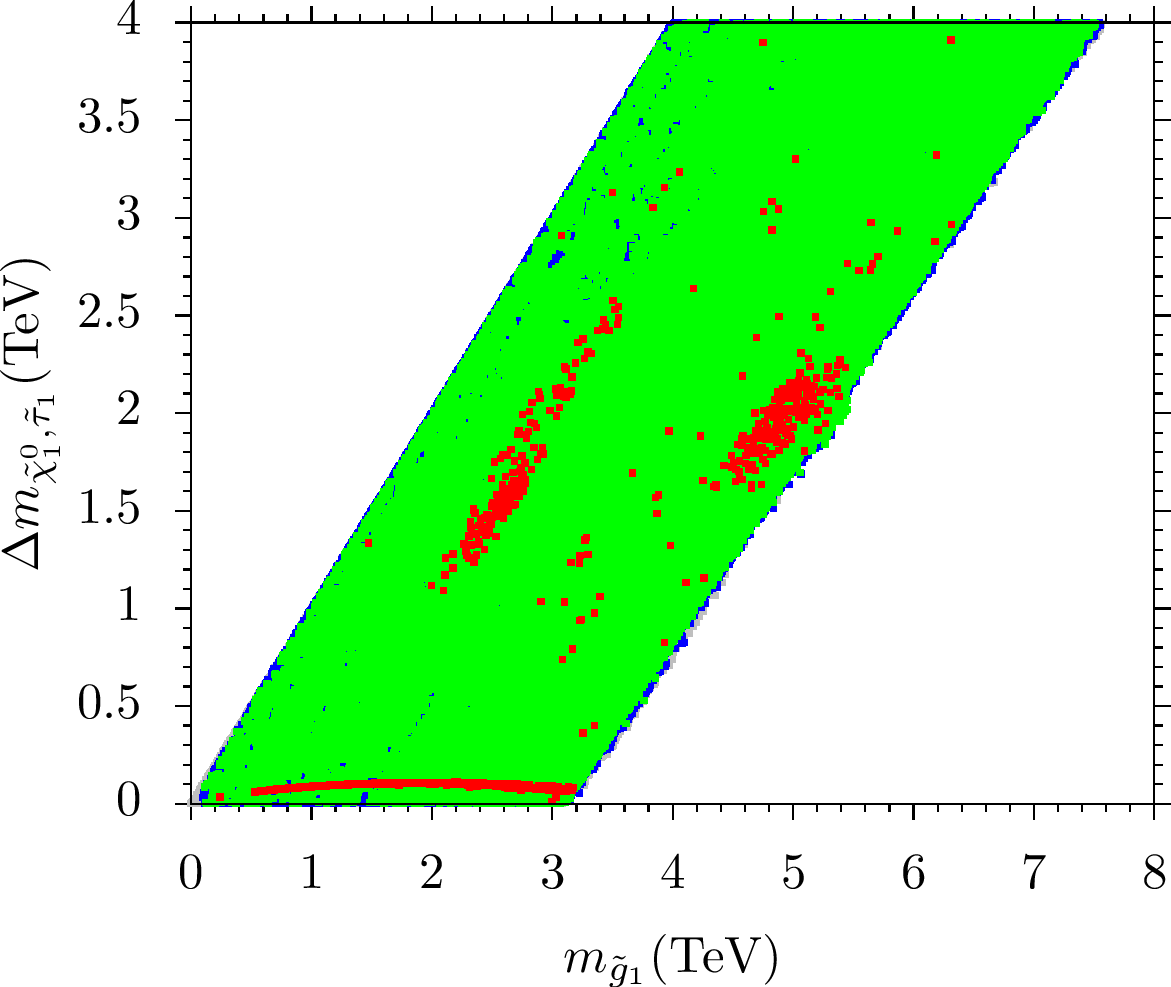}
\caption{Mass bounds and constraints in the $m_{\tilde g}-m_{\tilde \chi_{1}^{0}}$ and $m_{\tilde g}-\mid \Delta m_{\tilde \chi_{1}^{0},\tilde g}\mid$ planes with same color scheme as in Fig. \ref{fig3}.}
\label{fig4}
\end{figure}

\begin{figure}[h!]
\centering \includegraphics[width=8cm]{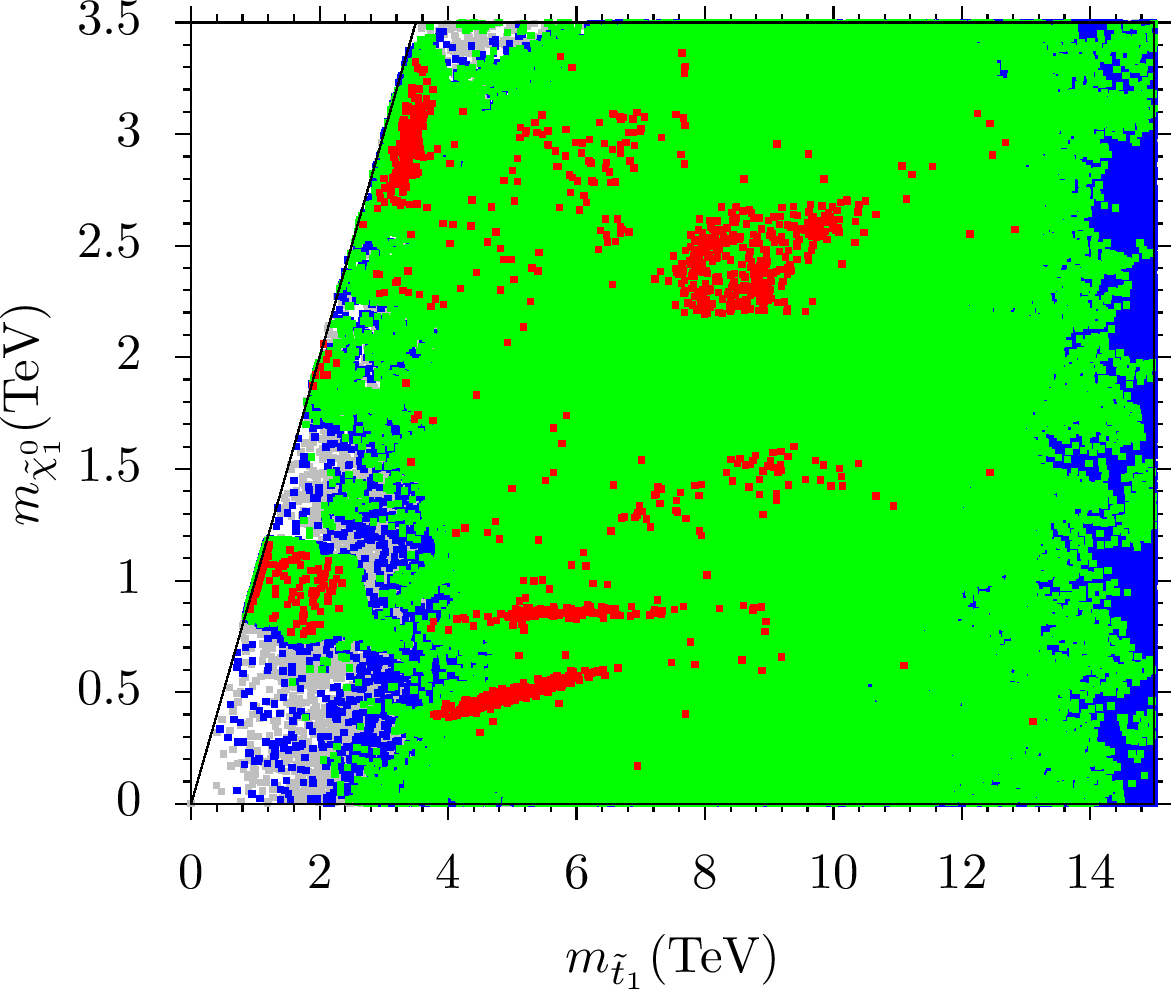}
\centering \includegraphics[width=8cm]{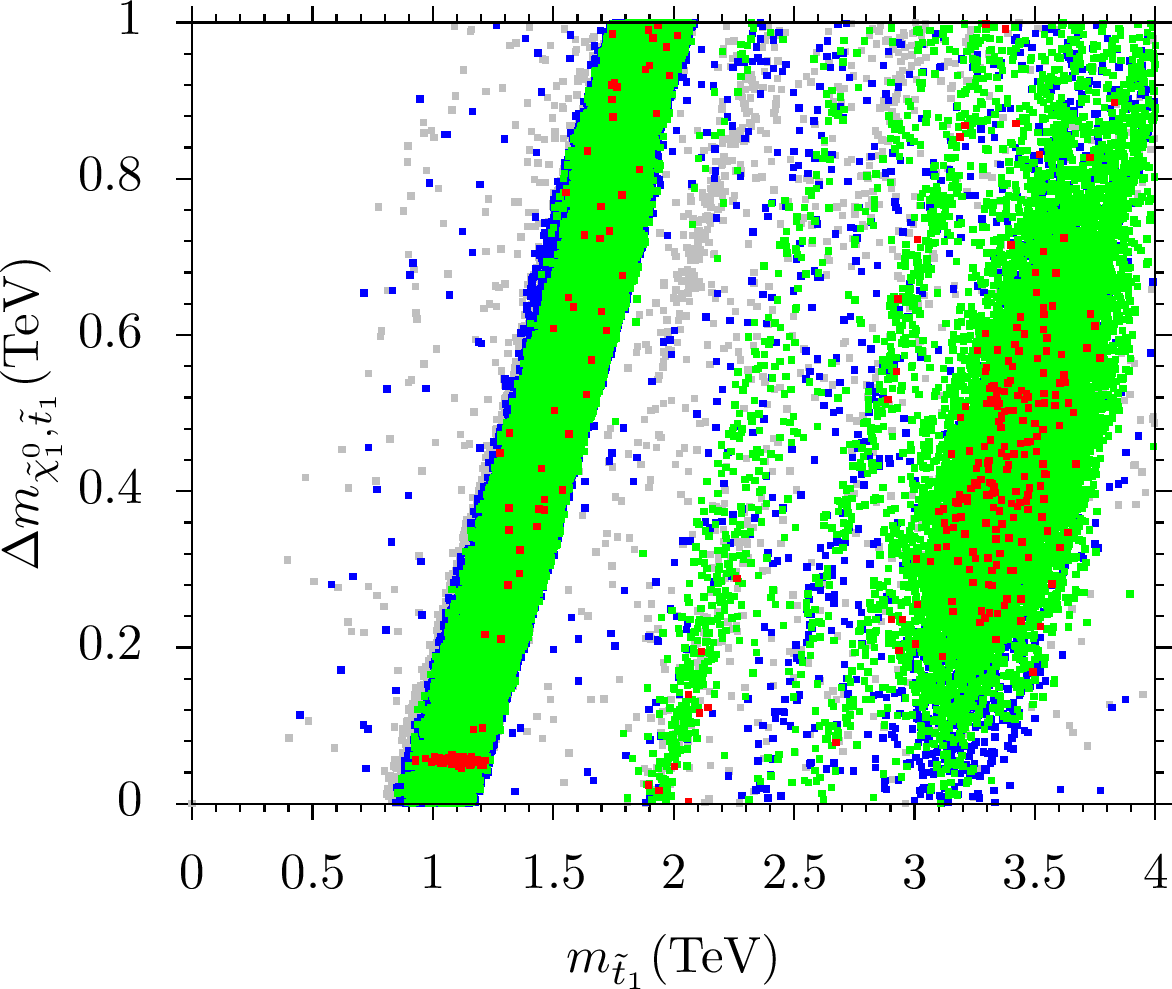}
\caption{Mass bounds and constraints in the $m_{\tilde t_1}$-$m_{\tilde \chi_{1}^{0}}$ and $m_{\tilde t_1}$-$\vert \Delta m_{\tilde \chi_{1}^{0},\tilde t_{1}}\vert$ planes with same color scheme as in Fig. \ref{fig3}.}
\label{fig5}
\end{figure}

\begin{figure}[h!]
\centering \includegraphics[width=8cm]{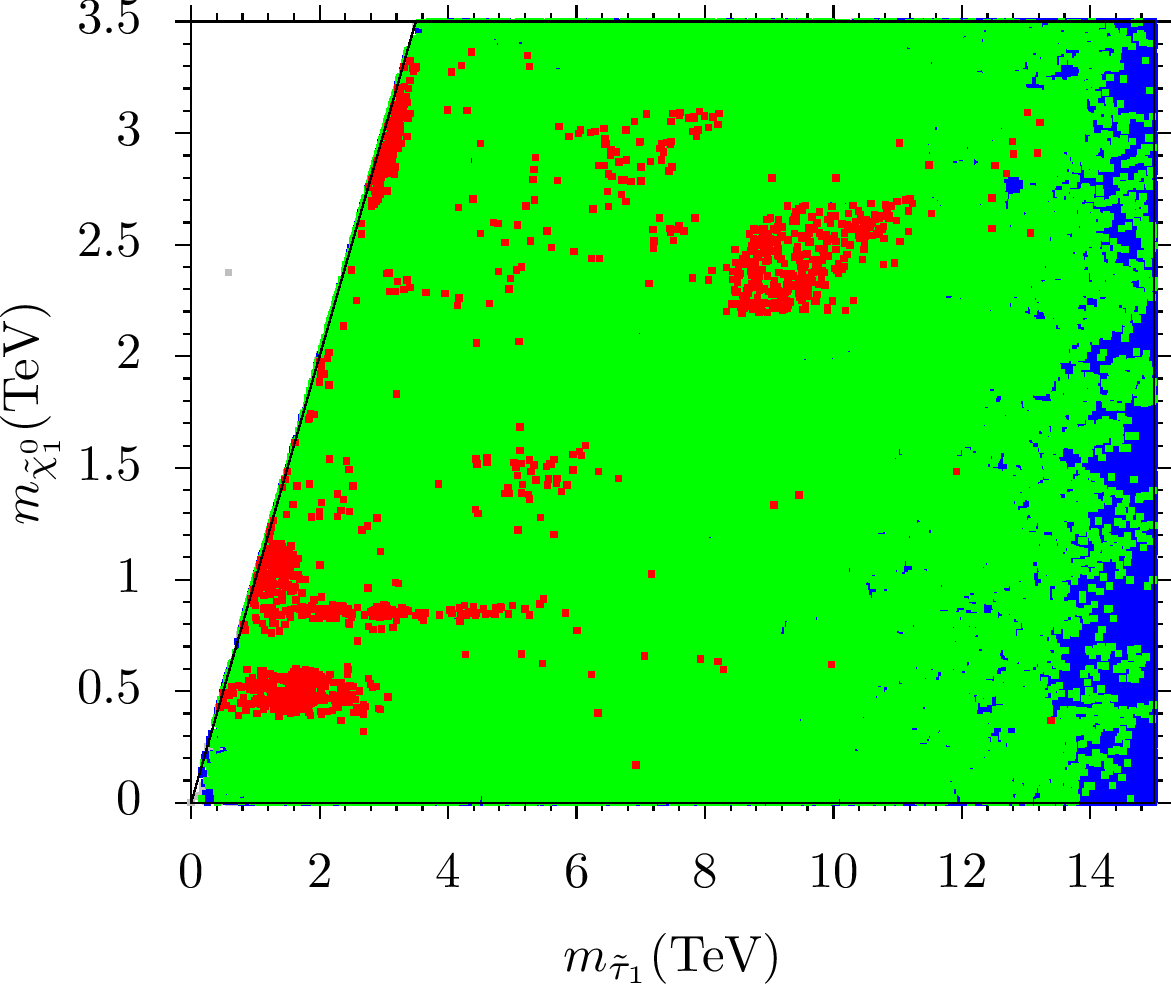}
\centering \includegraphics[width=8cm]{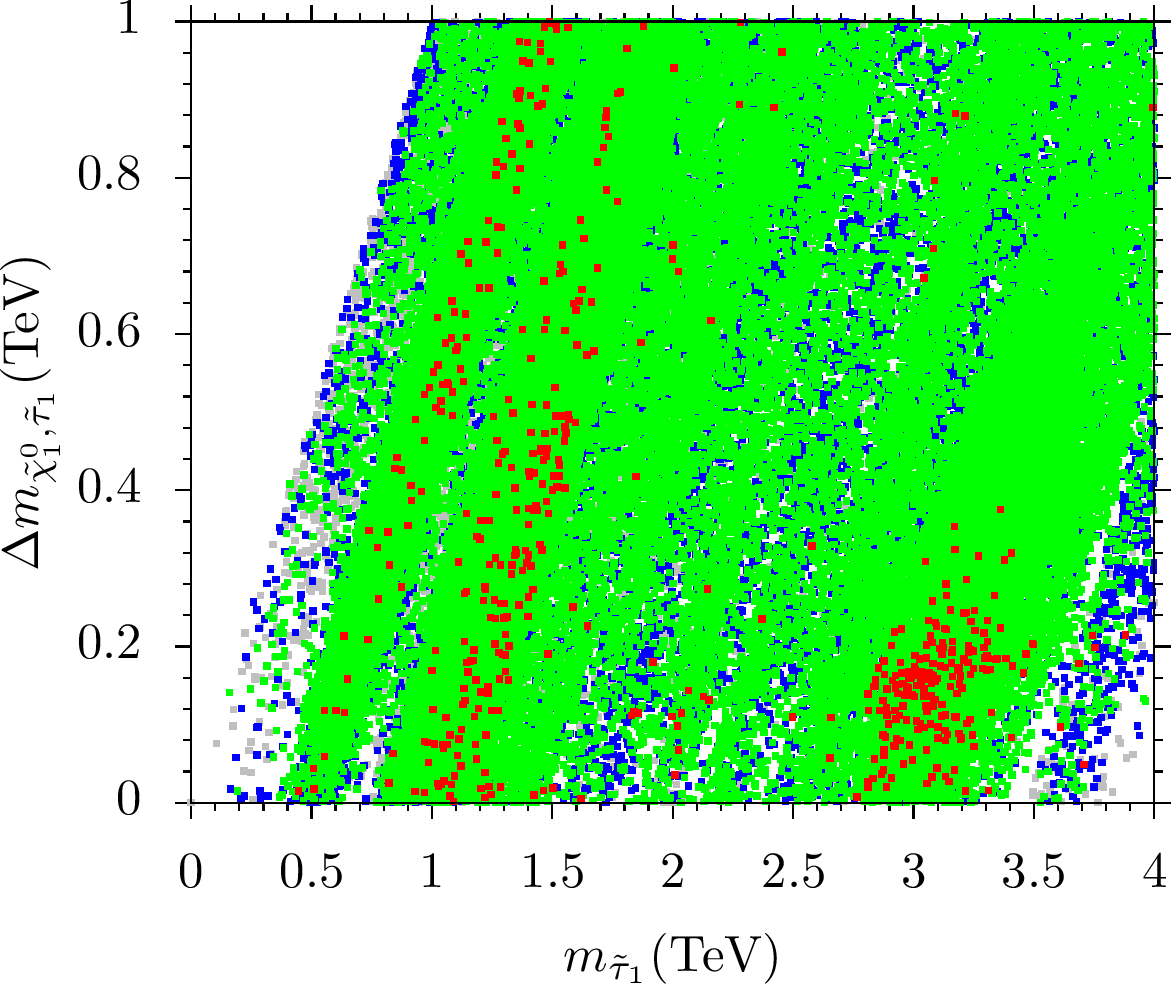}
\caption{Mass bounds and constraints in the $m_{\tilde \tau_1}$-$m_{\tilde \chi_{1}^{0}}$ and $m_{\tilde \tau_{1}}$-$\vert \Delta m_{\tilde \chi_{1}^{0},\tilde \tau_{1}}\vert$ planes with same color scheme as in Fig. \ref{fig3}.}
\label{fig6}
\end{figure}

\begin{figure}[h!]
	\centering \includegraphics[width=8cm]{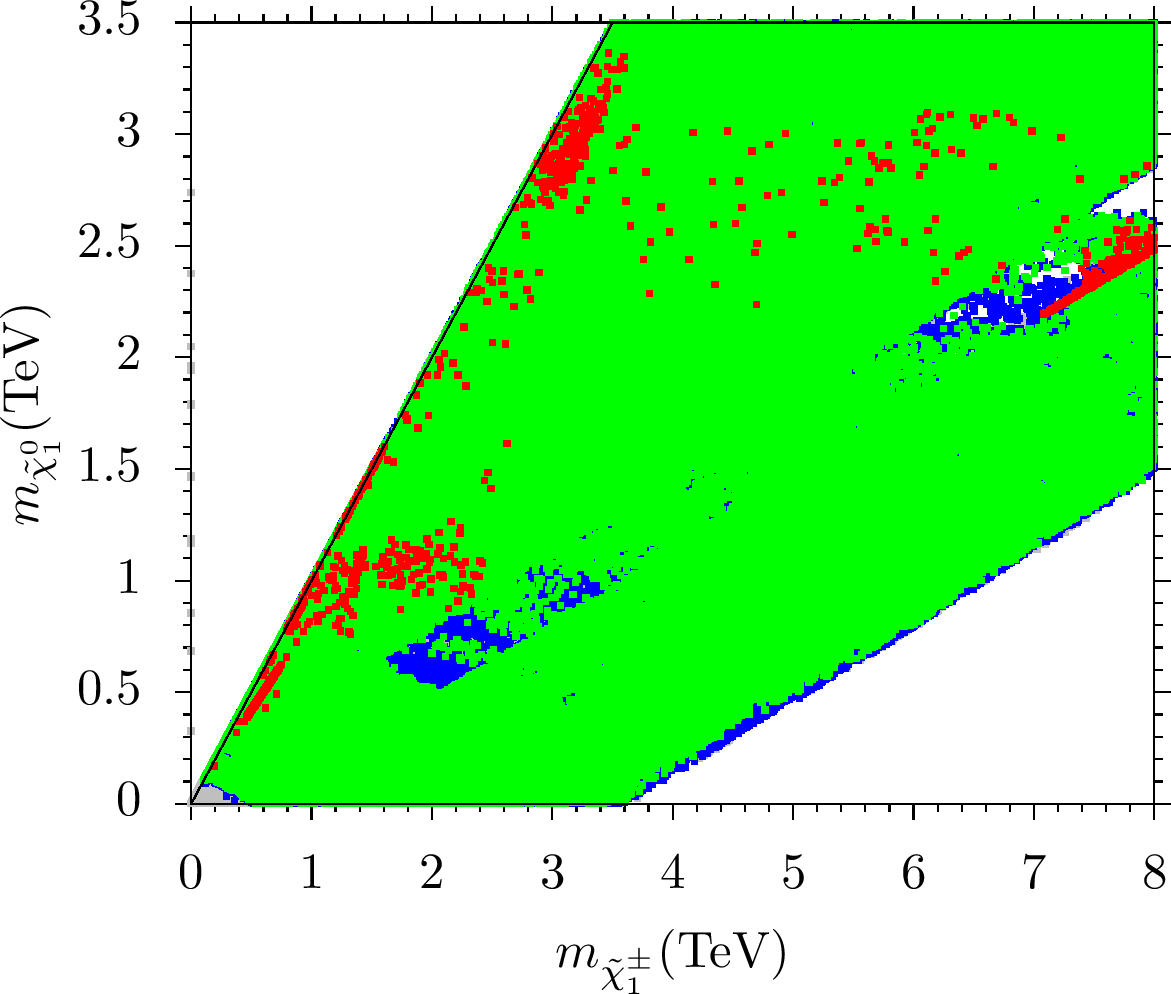}
	\centering \includegraphics[width=8cm]{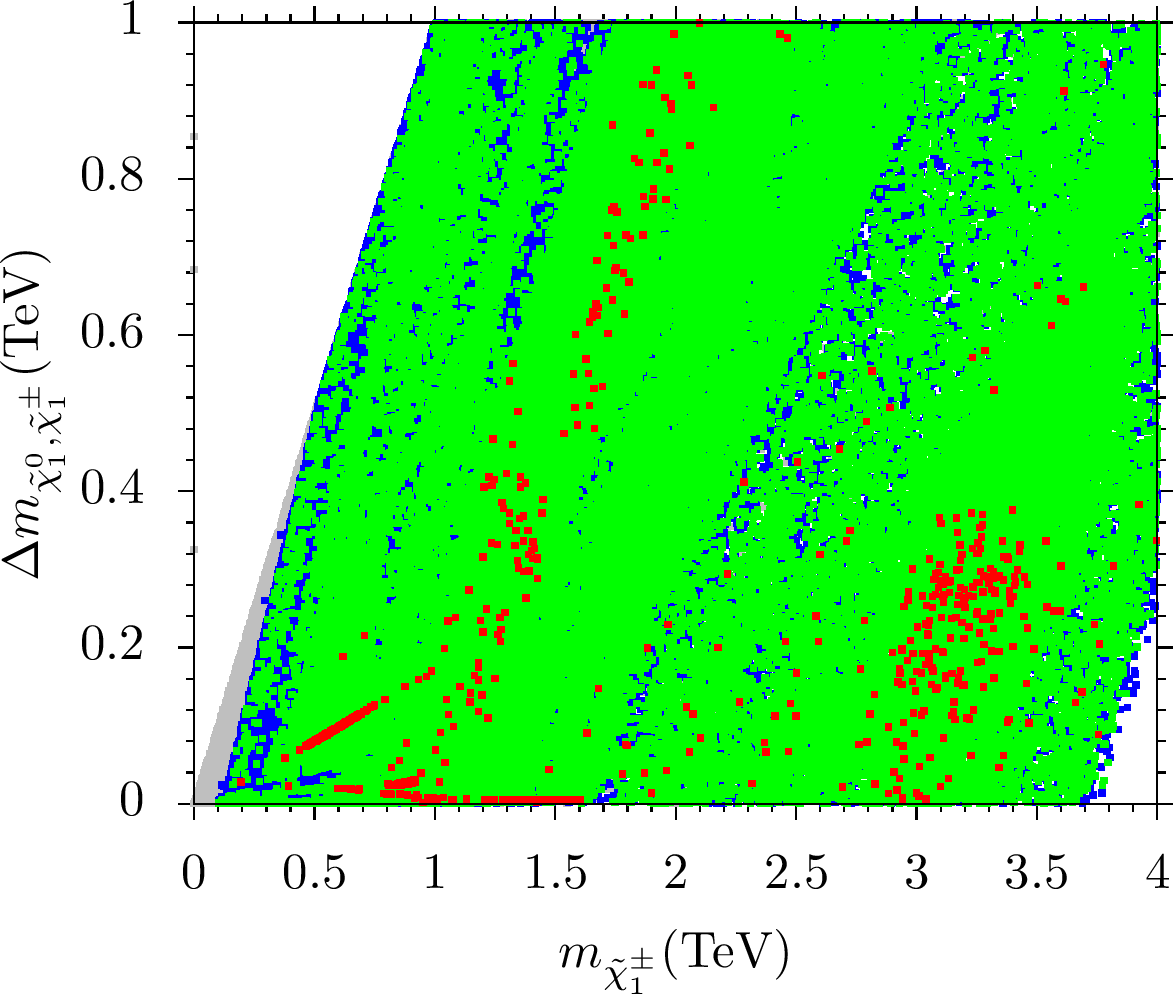}
	\caption{Plots in the $m_{\tilde \chi_{1}^{\pm}}-m_{\tilde \chi_{0}^{\pm}}$ and $m_{\tilde \chi_{1}^{\pm}}-\mid \Delta m_{\tilde \chi_{1}^{0},\tilde \chi_{1}^{\pm}}\mid$ planes.The color coding is the same as in Fig. \ref{fig3}.}
	\label{fig7}
\end{figure}

\begin{figure}[h!]
	\centering\includegraphics[width=8cm]{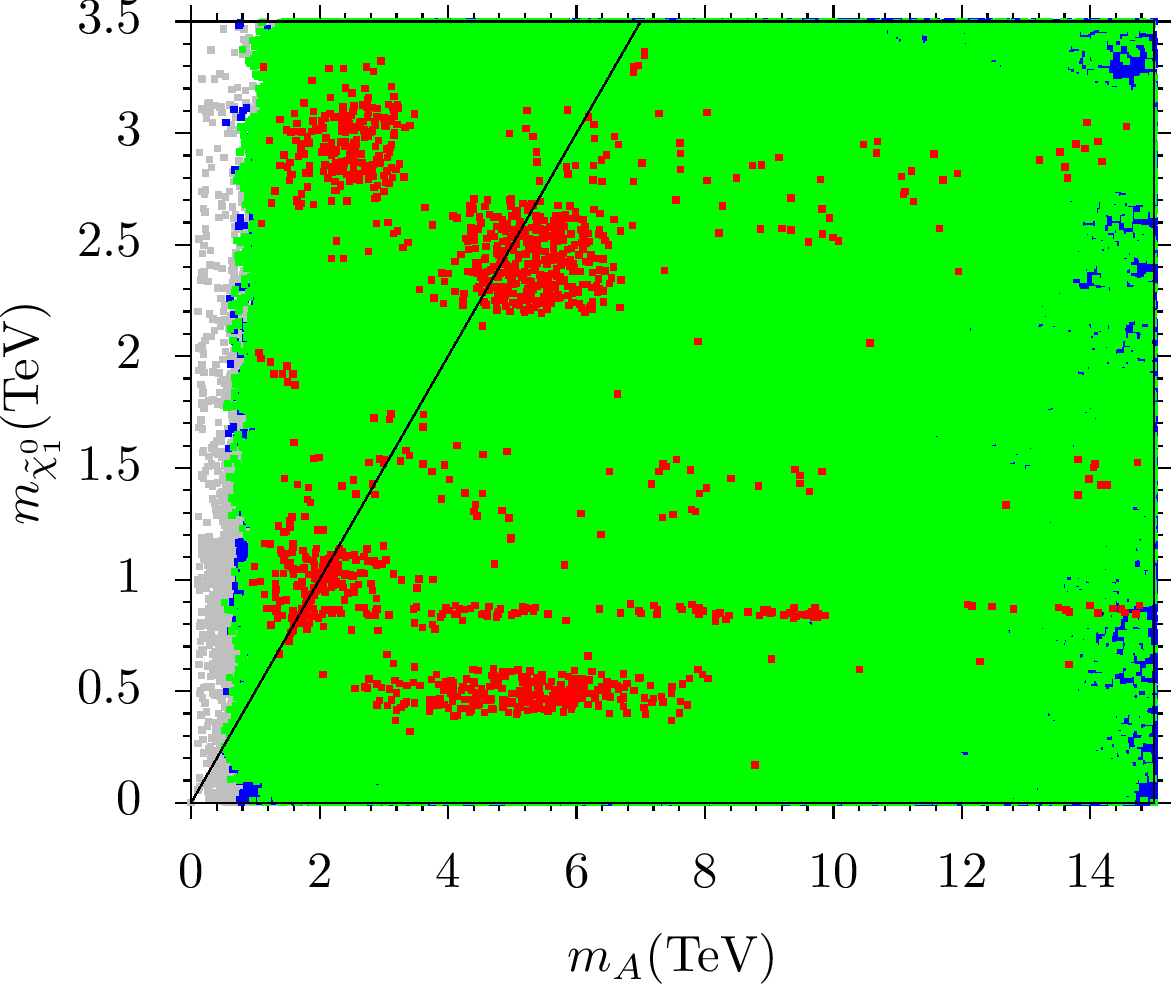}
	\centering\includegraphics[width=8cm]{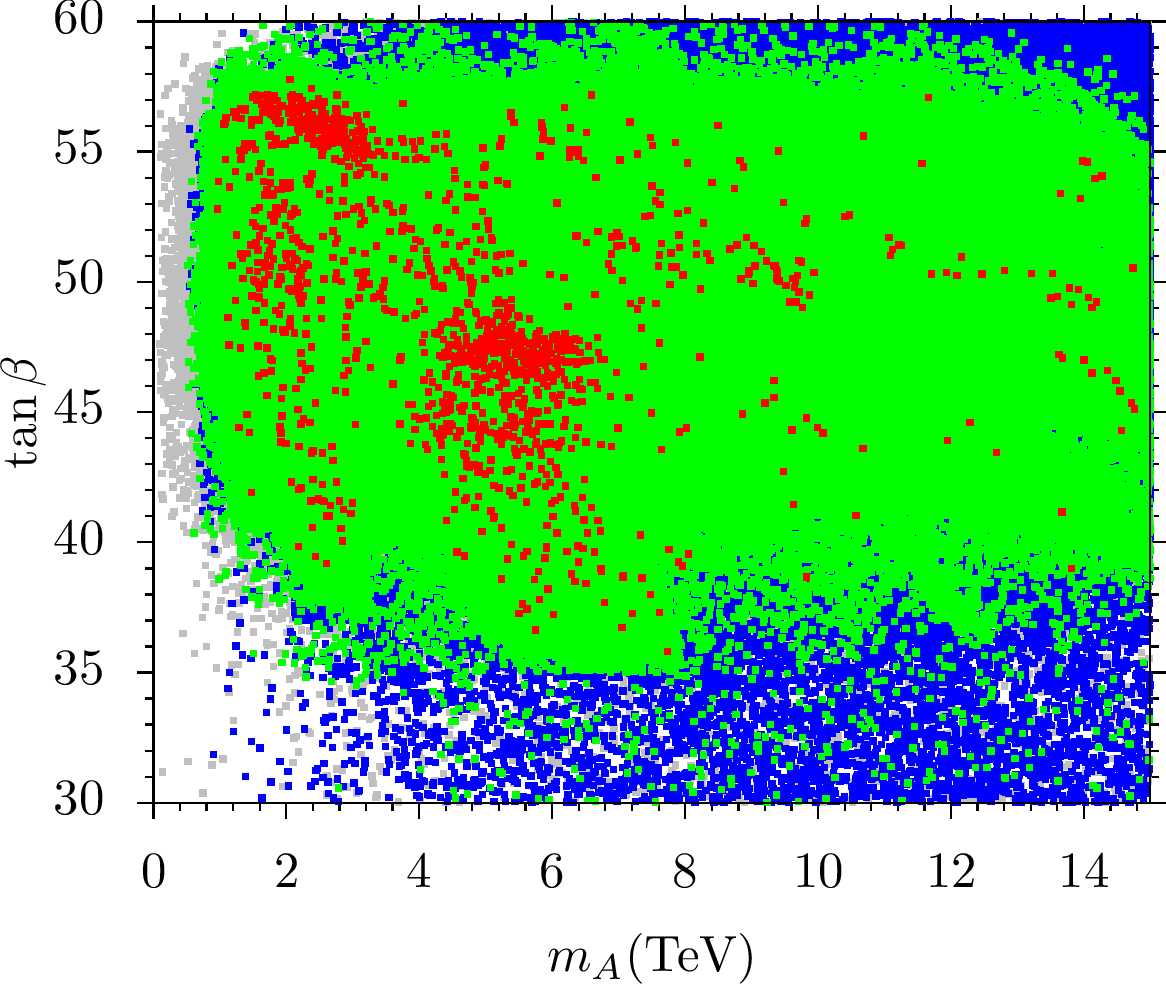}
	\caption{Plots in the $m_{A}-m_{\tilde \chi_{1}^{0}}$ and $m_{A}-\tan\beta$ planes.The color coding is the same as in Fig. \ref{fig3}}
	\label{fig8}
\end{figure}

In this section we will discuss the impact of $b-\tau$ YU on the parameter space of the fundamental parameters of SUSY $4-2-2$ model. In Figs. {\ref{fig1}-\ref{fig2}}, fundamental parameters are plotted versus $R_{b\tau}$. Gray points are consistent with the REWSB and neutralino LSP conditions. Blue points represent sparticle mass bounds, Higgs mass bound, B-physics bounds and  red points points satisfy $5\sigma$ Planck2018 bounds on the relic density of the LSP neutralino. The horizontal line shows the regions with $R_{b\tau}=1.1$, below which are the solutions with 10$\%$ or better $b-\tau$ YU.

In the top left panel of Fig.~\ref{fig1} we show plot in $m_{0}-R_{b\tau}$ plane. We see that as compare to $\mu > 0$ case where one needs heavy universal scalar mass parameter that is $ 7 \, \lesssim m_{0} \,\lesssim 20$ TeV ~\cite{Raza:2014upa}, we can have any value of $m_{0}$ between 0.5 TeV to 20 TeV for opposite sign gauginos with $\mu <0$ and $M_{2} < 0$ and $M_{3} > 0$ case consistent with 10$\%$ or better $b-\tau$ YU. This implies that we expect to have light to heavy spectrum with $R_{b\tau} \lesssim 1.1$. Similarly points consistent with relic density bounds (red points) can be between 1 TeV to 20 TeV. The concentration of red points at some places is the result of focused scans. In the right panel we display plot in $A_{0}/m_{0}-R_{b\tau}$ plane. We note that solutions (both blue and red) consistent with $R_{b\tau} \lesssim 1.1$ can be anywhere between $-3 \lesssim \, A_{0}/m_{0} \, \lesssim 3$. Here again we see that concentration of more red points for $A_{0}/m_{0}< 0$ is just because of more focused scans in this parameter space. The lower two panels of Fig. \ref{fig1} show the parameters $m_{H_d}$ (left) and $m_{H_u}$ (right) plotted against $R_{b\tau}$. For $m_{H_d}$, the entire range of our scan satisfies $10\%$ or better $t-b-\tau$ YU, whereas for $m_{H_u}$ the $b\tau$ unification condition is satisfied only for $m_{H_d}\lesssim 11$ TeV.

Fig. \ref{fig2} shows the parameters $M_2$, $M_3$ and $\tan\beta$ plotted against $R_{b\tau}$. 
In the top left panel we see that there is no preferred value of $M_{2}$ for $b\tau$ YU. We see that solutions consistent with $R_{b\tau} \lesssim 1.1$ can be anywhere between -10 TeV to 10 TeV. On the other hand plot in the top right corner shows that there is a grey region $0 \, \lesssim M_{3} \lesssim 1$ TeV excluded because of gluino mass bound. Except this grey region, solutions consistent with $R_{b\tau}$ YU can be realized from 1 TeV to 5 TeV. Plot in the lower panel displays that solutions satisfy 10$\%$ or better $b\tau$ YU requires $10 \, \lesssim \tan\beta \, \lesssim 60$.

\subsection{Sparticle Mass Spectrum Consistent with $b-\tau$ YU and Dark Matter Constraints}
In this section we display sparticle spectrum consistent with the $b-\tau$ YU, and other constraints discussed above including dark matter relic density bounds.

\noindent Fig. \ref{fig3} displays the NLSP sbottom mass $m_{\tilde b_{1}}$ plotted against the LSP neutralino mass $m_{\tilde \chi_{1}^{0}}$ in the left panel and their mass difference $\vert \Delta m_{\tilde \chi_{1}^{0},\tilde b_{1}}\vert$ versus $m_{\tilde \chi_{1}^{0}}$ in the right panel. Gray points satisfy the REWSB and neutralino as LSP conditions. Blue points satisfy the mass bounds and constraints from rare $B$-meson decays. Green points form a subset of blue points and satisfy $R_{b\tau} = 1.1$, whereas the red points are a subset of green points and are compatible with 5-$\sigma$ Planck2018 bounds on the relic density of the LSP neutralino. The diagonal line represents the co-annihilation region where the NLSP sbottom is mass degenerate with the LSP neutralino. The ref.~\cite{Gogoladze:2011ug} is the first study to show sbottom co-annihilation parameter space of $SU(5)$. Later on in ref.~\cite{Baer:2012by} it is shown that sbottom co-annihilation scenario is not compatible with $b-\tau$ YU with $SU(5)$ boundary conditions. It is important to note that in this article for the first time sbottom co-annihilation parameter space is presented consistent with $b-\tau$ YU and other constraints \footnote{As we mentioned before a couple of the NLSP sbottom solutions are also consistent with $t-b-\tau$ YU.}. To the best of our knowledge sbottom-neutralino co-annihilation parameter space consistent with $b-\tau$ YU has not been presented before in any GUTs model in general and in $4-2-2$ model in particular. The detailed study of this scenario will be presented elsewhere ~\cite{self}. In the left panel we see that the NLSP sbottom solutions compatible with dark matter relic density bounds(red points) are between 1 TeV to 3.4 TeV. Moreover, even if we relax the relic density constraint, we see that the NLSP sbottom solutions (green points) also have more or less same mass ranges. We also make a comment here that red points with the NLSP sbottom mass along the black line around 1 TeV or so but in this scenario instead of sbottom, chargino is the NLSP. But red points along the line with mass 2 TeV and above, sbottom becomes true NLSP and chargino becomes next to NLSP. Plot in the right panel shows mass difference between NLSP sbottom and LSP neutralino as a function of NSLP sbottom mass. We would like to remind readers that in this study we demand $\frac{\Delta m_{NLSP,LSP}}{m_{LSP}}\lesssim 10 \%$ where $\Delta m_{NLSP,LSP}= m_{NLSP}-m_{LSP} $. So the red points with small mass difference represent sbottom NLSP solution. We also comment here that if the mass difference is larger than $b$ quark mass, the available channel to search for NLSP sbottom is 
\begin{equation}
pp\rightarrow {\tilde b_{1}}{\tilde b_{1}^{*}X}\rightarrow b\bar{b} + \cancel{E}_{inv},
\end{equation}
where $\tilde b_{1}\rightarrow b \tilde \chi_{1}^{0}$. 

Moreover, there may also exist same sign sbottom pair productions $\tilde b_{1} \tilde b_{1}$ and ${\tilde b_{1}^{\ast}} {\tilde b_{1}^{\ast}}$. Recently there have been some searches for the light sbottom. For example the ATLAS collaboration have shown that for $\tilde b_{1}\rightarrow  b \chi_{2}^{0}\rightarrow \tilde b h \chi_{1}^{0}$ with $\Delta m_{\chi_{1}^{0},\chi_{2}^{0}}=$ 130 GeV sbottom mass can be ruled out up to the 1.5 TeV and 0.85 TeV \cite{ATLAS:2019gdh,ATLAS:2021pzz} respectively. Similarly, with $\Delta m_{\chi_{1}^{\pm},\chi_{1}^{0}}=$ 100 GeV, for $\tilde b_{1}\rightarrow  t \chi_{2}^{0}$ sbottom mass can be excluded up to 1.6 TeV \cite{ATLAS:2019fag}. Moreover for $\tilde b_{1}\rightarrow \tilde b \chi_{1}^{0}$ ($b-jets + \cancel E$) NLSP sbottom can be excluded up to 1.270 TeV for massless neutralino. In case of $m_{\tilde b_{1}}\approx m_{\tilde \chi_{1}^{0}}$, one may employ dedicated secondary-vertex identification techniques to exclude $m_{\tilde b_{1}}$ up to 660 GeV for $\Delta m_{{\tilde b_{1}},{\tilde \chi_{1}^{0}}} \sim $ 10 GeV~\cite{ATLAS:2021yij}. Similarly, for $\tilde b_{1}\rightarrow  b \chi_{1}^{0}$(monojet) sbottom mass can be excluded up to 600 GeV. Moreover, according to \cite{ATLAS:2021yij}, there are no mass limit on sbottom quark mass if it accedes to 800 GeV. As we can see that in the first two cases sbottom is not the NLSP but the last two channels are relevant. So our results are save. We hope that in future collider searches, these solutions will be accessible to Run-3.

Fig.~\ref{fig4} shows the LSP neutralino mass $m_{\tilde \chi_{1}^{0}}$ against the NLSP gluino mass $m_{\tilde g}$ (left) and and their mass difference $\vert \Delta m_{\tilde \chi_{1}^{0},\tilde g}\vert$ versus the NLSP gluino mass. Color coding is the same as in Fig.~\ref{fig3}, except we do not impose gluino mass bounds shows in ~\ref{sec:scan}. Here we see that red points along the diagonal line are between 0.2 TeV to 3.2 TeV which is a much better results as compare to ~\cite{Raza:2014upa} where the NLSP gluino mass was about 1 TeV. In reference \cite{Gomez:2020gav} the reported maximum NLSP gluino mass is 2.6 TeV. Since they imposed $t-b-\tau$ YU as compare to our case of $b\tau$ YU  which is a relaxed condition, so our gain in NLSP gluino mass is understandable.  In the right panel we show difference of gluino and neutralino mass ($\Delta m_{\tilde g,\tilde \chi_{1}^{0}}$) as a function of gluino mass. It is evident that the red points corresponding to the diagonal lines have $\Delta m_{\tilde g,\tilde \chi_{1}^{0}}$ less than 100 GeV. In fact the other red points visible in the plot correspond to points away from the diagonal lines. In this scenario the most dominant channel is $\tilde g \rightarrow b\bar{b}\chi_{1}^{0}$. In fact we one can choose other channels too but in our case this channel is important as it provides the track-jets. But other decay channels will be suppressed by the high background contamination at low jet-$p_T$. Since we have a very compressed final state which means that the quarks will not have enough energy to create tracks and hence the background will dominate for the case of light quarks while with the b-quark we have a secondary vertex and tracks that we still can reconstruct. In ref.~\cite{ATLAS:2018yhd,ATLAS:2021ilc} one can extract mass limit on gluino mass in case of gluino-neutralino mass degenerate case which is about 1.2 TeV. This shows that our results are consistent with present searches but some solutions have already been excluded. A detailed collider analysis is needed to explore this scenario consistent with LHC Run-3 and future colliders. 

In the left panel of Fig.~\ref{fig5} we show plot in $m_{\tilde t_{1}}-m_{\tilde \chi_{1}^{0}}$ plane. Color coding is same as Fig.~\ref{fig3}. It should be noted that NSLP stop solutions are present in $b-\tau$ YU scenarios but not in $t-b-\tau$ YU case. Here again the NLSP stops solutions(red points) consistent with 5$\sigma$ dark matter relic density bounds are along the black line. We see that in our present scans such NLSP stop solutions are spread in the interval of $0.9\, m_{\tilde t_{1}} \, 3.5$ TeV. We want to make a comment here that in ref.~\cite{Raza:2014upa} NSLP stop solutions are upto 0.8 TeV. In previous studies of $b-\tau$ YU ~\cite{Baer:2012by,Raza:2018jnh}, the heaviest NLSP stop mass achieved was about 3 TeV. \textcolor{blue}.{In our present study somehow we do not have large density of green points in this region, so in results no red points. This is an artifact of scanning. Had we done some more focused scans, we would have populated this region of parameter space with more points so would get NLSP stop solutions too}. Plot in the right panel is in $\Delta m_{\tilde t_{1},\tilde \chi_{1}^{0}}-m_{\tilde \chi_{1}^{0}}$ plane.  We note that as compared to previous studies, for the red solutions the difference between the NLSP light stop mass and the LSP neutralino can be as large as 300 GeV. It is important to note that this large mass difference corresponds to large stop and neutralino masses such that $\frac{\Delta m_{{NLSP},{LSP}}}{m_{LSP}}\lesssim 10 \%$ still satisfies. Such a large mass differences kinematically allow decay channels like $\tilde t_{1}\rightarrow t \tilde \chi_{1}^{0}$ along with three body decay $\tilde t_{1}\rightarrow W + b + \tilde \chi_{1}^{0}$ and four body decay $\tilde t_{1}\rightarrow f + f^{'} + b+ \tilde \chi_{1}^{0}$. On the other hand, for small mass gap case, above mentioned decay channels are not allowed kinematically but the loop induced two-body decay of NLSP stop, $\tilde t_{1}\rightarrow c \chi_{1}^{0}$, is generally dominant mode as compared to the four-body channel ~\cite{Hikasa:1987db,Muhlleitner:2011ww}. For previous studies see \cite{Raza:2014upa} and for recent LHC	studies see \cite{ATLAS:2017drc,ATLAS:2017eoo,ATLAS:2017www,ATLAS:2017bfj,ATLAS:2019zrq,ATLAS:2021kxv,ATLAS:2020dsf,ATLAS:2020xzu,ATLAS:2021hza}. In all of these studies, the maximum stop mass considered is 1.2 TeV as compared to our case the minimum stop mass allowed by all constraint(red points) is about 800 GeV. Even for small mass gap where $\tilde t_{1}\rightarrow t \tilde \chi_{1}^{0}$ dominates, stop mass upto 550 GeV has been excluded \cite{ATLAS:2021kxv}. It is evident the NLSP stop mass  that  we have shown here lies beyond these exclusion limits, but we hope that the future LHC searches will probe it.

Fig.\ref{fig6} shows various mass bounds and constraints in the $m_{\tilde \tau_1}$-$m_{\tilde \chi_{1}^{0}}$ (left) and $m_{\tilde \tau_1}$-$\vert \Delta m_{\tilde \chi_{1}^{0},\tilde \tau}\vert$ (right) planes with the same color scheme as in Fig.\ref{fig3}. The left panel of Fig. \ref{fig6} displays the stau-neutralino coannihilation, whereas the right panel shows the mass difference between stau and neutralino. It can be seen that in our scan the light stau, degenerate in mass with neutralino, lies in the range $0.45 ~ \text{GeV} \lesssim m_{\tilde \tau_1} \lesssim 3.8 ~ \text{TeV}$. We note that our results are consistent with the results reported in \cite{Raza:2018jnh,Gomez:2020gav}. Moreover, from ref.~\cite{CMS:2022rqk} we note that our solutions are also consistent with the study published by CMS with 137 $\rm fb^{-1}$ at 13 TeV. We hope some of the parameter space we present here will be probed in LHC Run-3 and beyond.
\begin{figure}[!htb]
	\centering\includegraphics[width=8cm]{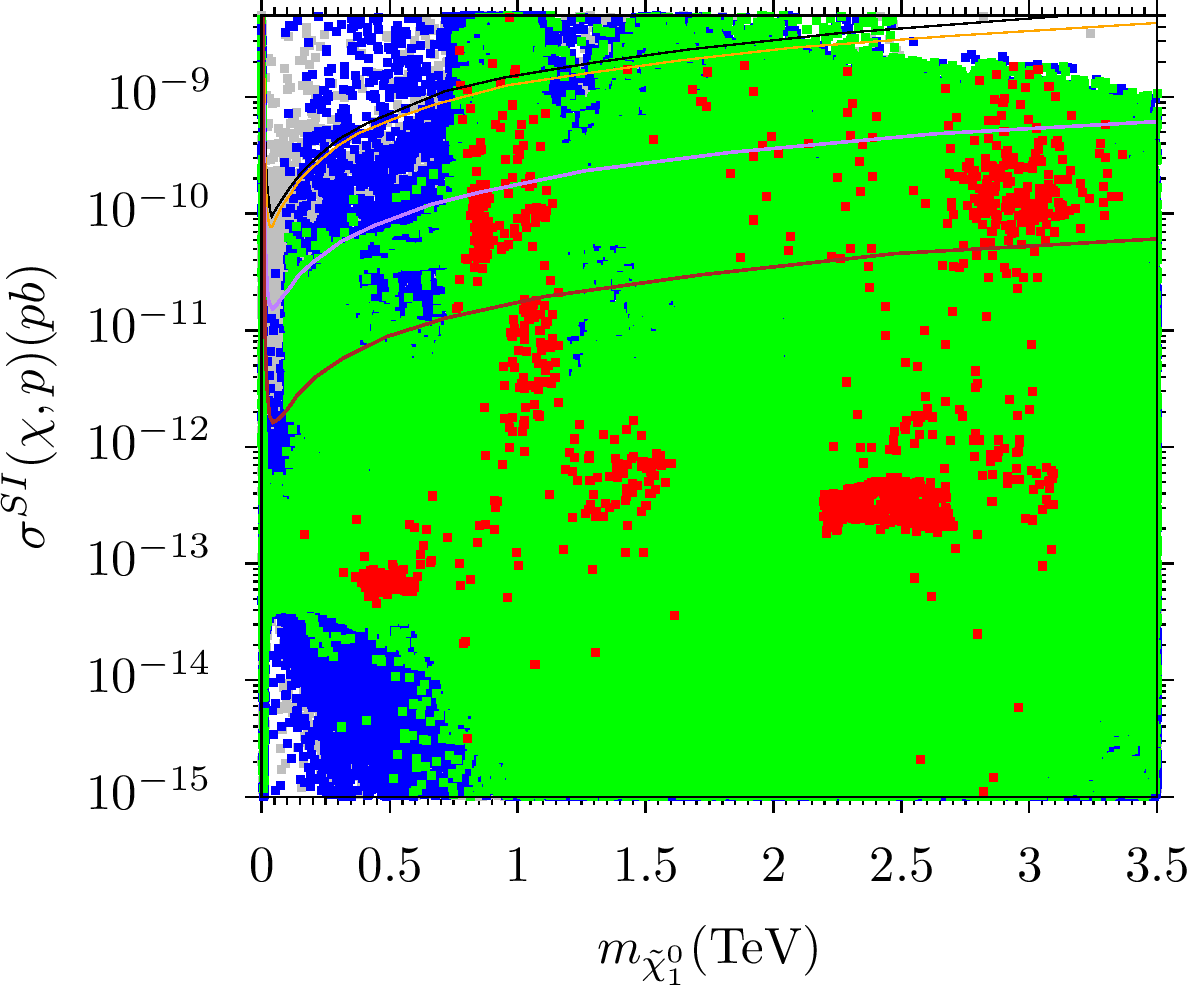}
	\centering\includegraphics[width=8cm]{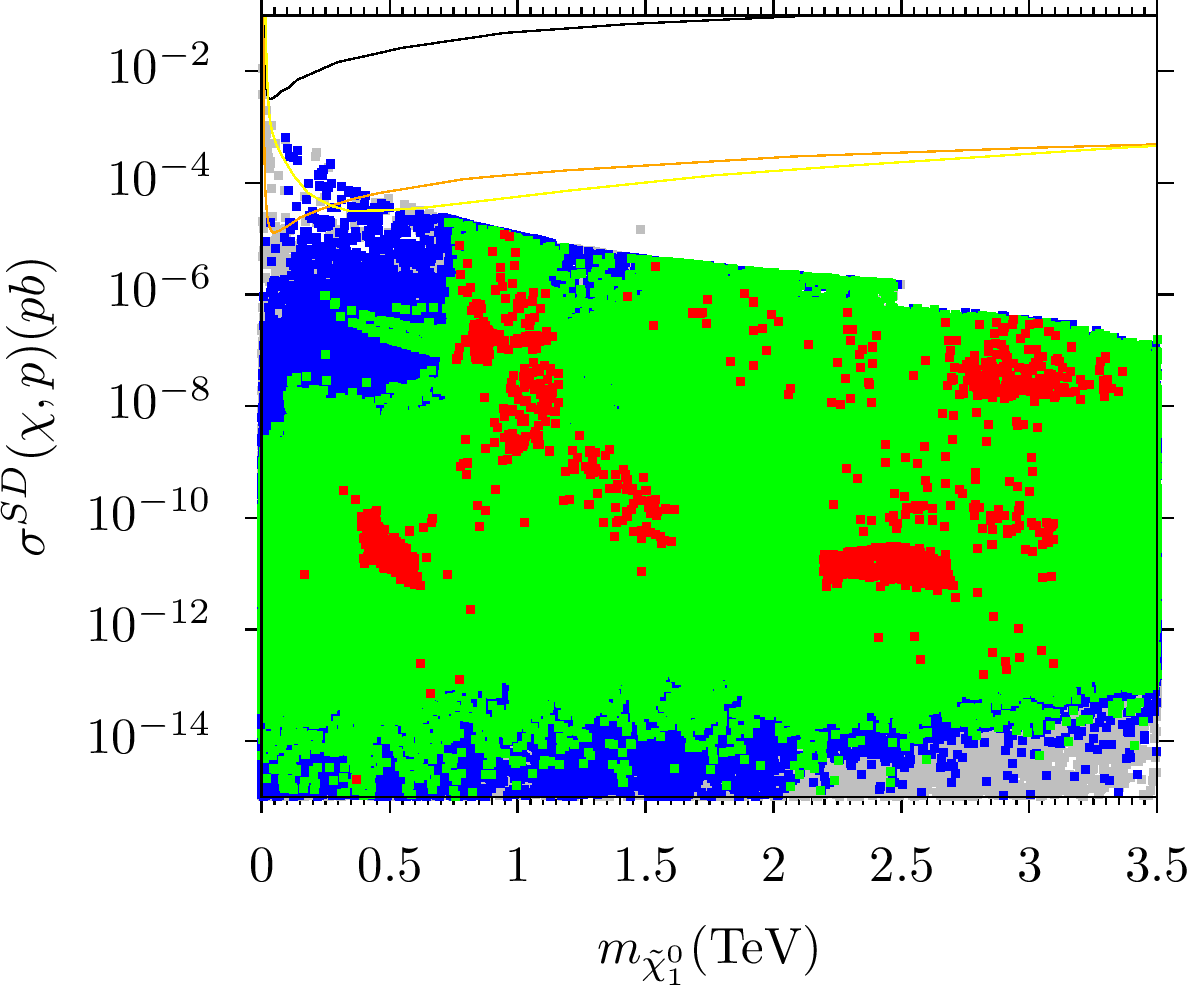}
	\caption{Plots in the $m_{\tilde \chi_{1}^{0}}-\sigma^{SI}(\chi,p)(pb)$ and $m_{\tilde \chi_{1}^{0}}-\sigma^{SD}(\chi,p)(pb)$ planes.The color coding is the same as in Fig. \ref{fig3}. Plots in the $m_{\tilde{\chi}_{1}^{0}}-\sigma^{{\rm SI}}$ and $m_{\tilde{\chi}_{1}^{0}}-\sigma^{{\rm SD}}$ planes (see text for the description of the bounds).}
	\label{fig9}
\end{figure}

In addition to the co-annihilation channels discussed above, our scans also yield charginio-neutralino coannihilation as shown in Fig. \ref{fig7}, where several constraints are displayed in the $m_{\tilde \chi_{1}^{\pm}}$-$m_{\tilde \chi_{1}^0}$ and $m_{\tilde \chi_{1}^{\pm}}$-$\vert \Delta m_{\tilde \chi_{1}^{0},\tilde \chi_{1}^{\pm}}\vert$ planes. It can be seen red points where chargino is degenerate in mass with the LSP neutralino is also consistent with $b-\tau$ YU in the mass range $0.2$ TeV $\lesssim m_{\tilde \chi_{1}^{\pm}} \lesssim 3.5$ TeV. Our results are consistent with \cite{Gomez:2020gav}. Moreover if we look at the recent searches for charginos, we note that for sleptons and as well as SM-boson mediated decays of $\tilde \chi_{1}^{+}\tilde \chi_{1}^{+}$ and $\tilde \chi_{1}^{\pm}\tilde \chi_{2}^{0}$ the 95$\%$ exclusion limits are given in \cite{ATLAS:2021ilc}. From this reference we can see that charginos degenerate with the LSP neutralino, any solutions heavier than 300 GeV are save. On the other hand in the parameter space where slepton masses are heavier than charginos, these slepton mediated decays will not take place. Since we also have heavier NLSP chargino solutions, we hope that such solutions will be probed in future LHC searches.

Besides the co-annihilation channels we also have Higgs resonance scenario where a pair of LSP neutralinos decay via CP-odd(even) higss A(H,h) to the SM particles. This may help in achieving the relic density in the allowed range. Fig.~\ref{fig6} shows that it is possible to have  solutions with $m_A \approx 2m_{\tilde \chi_{1}^{0}}$. We also note that in this scenario $m_{A}\sim m_{H}$. In the ref.~\cite{CMS:2022goy} it is shown that for $A$, $H \rightarrow \tau {\bar \tau}$, $m_{A} \lesssim$ 1.7 TeV is excluded for $\tan\beta\lesssim$ 30. Similarly, it is reported that for $\tan\beta\lesssim$ 10  $m_{A}\sim$ can be excluded for the values 1 TeV, 1.1 TeV and 1.4 TeV at Run 2, Run 3 and HL-LHC respectively  \cite{Baer:2022qqr,Baer:2022smj}.  From our plots we see that the range $A$-resonance solutions is between 0.4 Te to 3.5 TeV. So some part of the parameter space has already been explored by the LHC searches.

\subsection{Dark Matter Implications}
\label{sec:DM}

Finally, in this section we study the implications of $b-\tau$ YU and DM current and future searchers on the parameter space of $4-2-2$. We note the co-annihilation and the resonance scenarios we have discussed above the LSP is bino-type.

In Fig.~\ref{fig9} we show spin-independent (SI) scattering cross section (left) and spin-dependent (SD) scattering cross section (right) of nucleons-neutralino as functions of the LSP neutralino mass. In the left panel, solid black and yellow lines respectively represent the current LUX \cite{Akerib:2016vxi} and XENON1T \cite{Aprile:2017iyp} bounds, and the blue and brown lines depict the projection of future limits \cite{Aprile:2015uzo} of XENON1T with 2 $t\cdot y$ exposure and XENONnT with 20 $t\cdot y$ exposure, respectively. In the right plot, the black solid line is the current LUX bound \cite{Akerib:2017kat}, the orange line shows the future LZ bound \cite{Akerib:2016lao} and yellow line represent the IceCube DeepCore.ref[].

Plot in the $m_{\tilde{\chi}_{1}^{0}}-\sigma_{SI}$ plane shows that almost all red solutions are below the  But we see that only except handful of red points having mass around 1 TeV, all other red points are below black line and yellow lines. But some of the red points are accessible to future XENON1T with 2 $t\cdot y$ (dashed blue line) and nearly half of the red solutions can be probed by XENONnT with 20 $t\cdot y$ exposure (dashed brown line).  This scenario where we see that red solutions have relatively small neutralino-nucleon spin-independent scattering cross-sections suggest that LSP neutralino dominantly of bino-type. The plot in the $m_{\tilde{\chi}_{1}^{0}}-\sigma_{SD}$ plane shows that our solutions are consistent with the current and future reaches of the direct-detection experiments.

\begin{table}[h!]
	\centering
	\scalebox{0.7}{
		\begin{tabular}{|l|c|c|c|c|c|c|}
			\hline
			\hline
			&  Point 1 & Point 2  &  Point 3  & Point 4 & Point 5 & Point 6  \\
			\hline
			$m_{0}$        &  4979   & 6679   & 3426    & 2046 &  2357 & 2524   \\
			$M_{2} $          & 7491   & 7293    & 972.6  & 2535   & 2992 & 3049    \\
			$M_{3}$         &  1347  & 945.3   & 3033   & 1563  & 1002 & 1183\\
			$A_{0}/m_{0}$  &  -0.4446 & -0.9796 & -0.8262   & -1.243  & -1.842  & 1.638 \\
			$\tan\beta$       & 55.5    & 55.1    & 52.15  &  54.77   & 46.5 & 49.73\\
			$m_{H_d}$         & 5641   & 5979   & 4856   &  3333  & 2759  & 3652\\
			$m_{H_u}$          & 337.4   & 1319    & 595.5  & 2683  & 2152 & 2774\\
			\hline
			$m_h$                & {\bf 124}   & {\bf 125}    & {\bf 123}   & {\bf 123}  & {\bf 125} & {\bf 125}\\
			$m_H$                 & 3741 & 3925  & 3731  & 1318.28 & 1683.3 & 2116  \\
			$m_A$                 & 3716 & 3899  & 3731  & 1309  & 1672  & 2102    \\
			$m_{H^{\pm}}$         & 3742 & 3926  & 3757 & 1322 & 1686  & 2118\\
			\hline
			$m_{\tilde{\chi}^0_{1,2}}$
			&  {\color{red}2342}, 2668 & {\color{red}2236},  4805 & {\color{red} 800}, 812 & {\color{red} 955},  1150& {\color{red} 990}, 1830 & 1037,1424   \\
			$m_{\tilde{\chi}^0_{3,4}}$
			& 2670, 6284 & 4807, 6183 & 4283, 4283  & 1150, 2109.  & 1835, 2503 & 1426, 2547  \\
			
			$m_{\tilde{\chi}^{\pm}_{1,2}}$
			&  2582, 6240 &  4696, 6118& {\color{red} 813}, 4250  & 1105, 2079  & 1801, 2486  & 1386, 2525 \\
			\hline
			$m_{\tilde{g}}$ &  3081   &{\color{red} 2329 }& 6299.6  & 3378 & 2285 &2652         \\
			$m_{ \tilde{u}_{L,R}}$
			& 7146, 5524 & 8178,6868 & 6343, 6277  & 3837, 3532  & 3519, 3029 & 3821, 3358   \\
			$m_{\tilde{t}_{1,2}}$
			& 3210, 5412 & 3929, 5975 & 5009, 5174 & 2165, 2741  & {\color{red}1043}, 2318 & 1361,2507   \\
			\hline $m_{ \tilde{d}_{L,R}}$
			&7147, 5565 & 8179, 6919 & 6343, 6362  & 3837, 3533 & 3520, 3021  &3822, 3358    \\
			$m_{\tilde{b}_{1,2}}$
			& {\color{red}2554}, 5427 & 3821, 5968 & 4863, 5075  & 2226, 2728  & 1581, 2345  & 1803, 2535 \\
			\hline
			$m_{\tilde{\nu}_{e,\mu}}$
			& 6811   & 8055 & 3401& 2607 & 3036  & 3178     \\
			$m_{\tilde{\nu}_{\tau}}$   & 6126 & 7169 & 2847  & 2188 & 2669 & 2713 \\
			\hline
			$m_{ \tilde{e}_{L,R}}$
			& 6806, 5444 & 8051, 7011 & 3402, 3643  & 2608, 2234  & 3034, 2520 & 3177, 2712   \\
			$m_{\tilde{\tau}_{1,2}}$
			&  3360, 6106 & 4646, 7151 & 2464, 2848  & {\color{red}968}, 2188  & 1434, 2662  &1328, 2707 \\
			\hline
			
			$\sigma_{SI}({\rm pb})$
			&  $1.55\times 10^{-9}$& 1.0$\times10^{-12}$ & 3.14$\times 10^{-16}$ & 2.93$\times 10^{-10}$ & $5.6\times 10^{-12}$ & $7.66\times 10^{-10}$\\

			$\sigma_{SD}({\rm pb})$
			&  4.8$\times 10^{-8}$ &1.75$\times 10^{-10}$ &5.9$\times 10^{-10}$   &8.3$\times 10^{-7}$  & $1.4 \times 10^{-8}$ & $1.39 \times 10^{-7} $\\
			$\Omega_{CDM}h^{2}$
			&  0.120 & 0.1207 &0.126  & 0.120  & 0.1175 & 0.124\\
			\hline
			$R_{b\tau}$    & 1.01 & 1.01062  &1.001  & 1.0087 & 1.012 & 1.08\\
			\hline
			\hline
		\end{tabular}
	}
	\caption{Fundamental parameters and resulting sparticle mass spectrum are shown. All masses are given GeV. }
	\label{table1}
\end{table}

Finally, we also show six benchmark points in Table \ref{table1} which summarize our findings for co-annihilation scenarios. Point 1 displays an example of NLSP sbottom. Here we see that NLSP sbottom is about 2.554 TeV with LSP neutralino which is a bino of mass about 2.342 TeV and $b\tau$ YU is about 1$\%$. We also note that in this case the $BR({\tilde b_{1}} \rightarrow b \tilde \chi_{1}^{0})$ is 100$\%$. Point 2 represents the NLSP gluino scenario. In this point gluino mass is about 2.329 TeV and the LSP neutralino, which is a bino, mass is about 2.336 TeV. Moreover, here $R_{b\tau}=1.01$ and  $BR({\tilde g \rightarrow b {\bar b}{\tilde \chi_{1}^{0}}})=$0.6645 and $BR({\tilde g \rightarrow c {\bar c}{\tilde \chi_{1}^{0}}})=$0.1324. Point 3 depicts stop-neutralino co-annihilation scenario. Here NLSP stop mass is about 1.042 TeV and LSP neutralino (bino) mass is about 0.990 TeV, $R_{b\tau}=1.01$  and $BR({\tilde t_{1}} \rightarrow c \tilde \chi_{1}^{0})$ is 100$\%$. Point 4 is an example of chargino-neutralino coannihilation where $m_{\tilde \chi_{1}^{0}}=$0.813 TeV and LSP neutralino which is dominantly a bino with admixture of wino, has mass around 0.8 TeV. Here $R_{b\tau}=$1.00 and $BR({\tilde \chi_{1}^{\pm} \rightarrow q_{i} {\bar q_{i}}{\tilde \chi_{1}^{0}}})$ is about 33$\%$ where $i=u,d$ quarks and $BR({\tilde \chi_{1}^{\pm} \rightarrow l_{i} {\bar l_{i}}{\tilde \chi_{1}^{0}}})$ is about 11$\%$ where $i=e,{\mu},{\tau}$ leptons. Similarly points represents stau-neutralino co-annihilation case. Here we see that NLSP stau mass is about 1.042 TeV and LSP neutralino mass is about 0.990 TeV. Moreover, this is an example of 100$\%$ ${b\tau}$ YU with $BR({\tilde \tau_{1}} \rightarrow \tau \tilde \chi_{1}^{0})=$1.  It can be seen that in point 6 $m_A$ and $m_H$ are almost degenerate, so we can regard them either $A$ or $H$-resonance solution. We note that for this point LSP bino mass is about 1.037 TeV and $m_{A}=$2.012 TeV and $m_{H}=$2.116 TeV. Moreover,  the dominant branching fraction is $BR({A/H} \rightarrow b \bar{b})=$0.8582 and sub-dominant branching fraction $BR({A/H} \rightarrow \tau \bar{\tau})=$0.1352 $R_{b\tau}=$1.08.

\section{Conclusion}
\label{sec:conc}
In this article we revisit the $b$-$\tau$ YU in SUSY 4-2-2 model. We present for the first time sbottom-neutralino co-annihilation scenario consistent with  $b$-$\tau$ YU and known experimental collider and astrophysical bounds. In addition to it, we have also shown gluino-neutralino, stop-neutralino, stau-neutralino, chargino-neutralino and A-resonance scenarios. We show that all such solutions are consistent with existing experimental collider constraints, Planck2018 dark matter relic density bounds and as well as direct and indirect bounds on neutralino-nucleons scattering cross sections. We further show that in sbottom-neutralino co-annihilation scenario, sbottom mass is about 2 TeV, whereas in the case of gluino-neutralino and stop-neutralino, gluino mass can be in the range between 1 TeV to 3 TeV and stop mass is in the range of 1 TeV to 3.5 TeV. Stau and chargino masses can also be as heavy as 3.5 TeV in case of co-annihilation scenario. Similarly A-resonance solutions are also in the range of 0.5 TeV to 3.5 TeV. We anticipate that some part of the parameter space will be assessable in the supersymmetry searches in LHC Run-3 and future runs.

\section{Acknowledgement}
The work of S.N  and  M. B is supported by the United Arab Emirates University (UAEU) under UPAR Grants No. 12S093. SR thanks Qaisar Shafi to introduce YU scenario in the SUSY 4-2-2 model.

\end{document}